\documentclass[nofootinbib,prd,aps,onecolumn,preprintnumbers,amsmath,amssymb,superscriptaddress]{revtex4}
\usepackage{amsmath}
\usepackage{graphicx}
\usepackage{float}
\usepackage{dcolumn}
\usepackage{appendix}
\usepackage{bm}
\usepackage{amssymb}
\usepackage{latexsym}
\usepackage{subfigure} 
\usepackage{epstopdf}
\usepackage{threeparttable}
\usepackage{booktabs}
\usepackage{tabularx}
\usepackage{mathrsfs}
\usepackage[normalem]{ulem}        
\usepackage{cancel}
\usepackage{hhline}
\usepackage{multirow}
\usepackage{color}
\usepackage[colorlinks,linkcolor=magenta,anchorcolor=blue,citecolor=green]{hyperref}

\usepackage{pifont}
\usepackage{makecell}
\def\be{\begin{equation}}
\def\ee{\end{equation}}
\def\bea{\begin{eqnarray}}
\def\eea{\end{eqnarray}}

\begin{document}
\title{ADM formulation and Hamiltonian analysis of $f(Q)$ gravity}

\author{Kun Hu}
\email{hukun@mails.ccnu.edu.cn}
\affiliation{Institute of Astrophysics, Central China Normal University, Wuhan 430079, China}

\author{Taishi Katsuragawa}
\email{taishi@mail.ccnu.edu.cn (co-corresponding author)}
\affiliation{Institute of Astrophysics, Central China Normal University, Wuhan 430079, China}

\author{Taotao Qiu}
\email{qiutt@hust.edu.cn (corresponding author)}
\affiliation{School of Physics, Huazhong University of Science and Technology, Wuhan, 430074, China}

\begin{abstract}
$f(Q)$ gravity is an extension of the symmetric teleparallel equivalent to general relativity.
We demonstrate the Hamiltonian analysis of $f(Q)$ gravity with fixing the coincident gauge condition. 
Using the standard Dirac-Bergmann algorithm, we show that $f(Q)$ gravity has 8 physical degrees of freedom. 
This result reflects that the diffeomorphism symmetry of $f(Q)$ gravity is completely broken due to the gauge fixing.
Moreover, in terms of the perturbations, we discuss the possible mode decomposition of these degrees of freedom.

\end{abstract}

\maketitle

\section{Introduction}

General relativity (GR) is a well-known classical theory for describing gravity, where Einstein starts with a few convincing physical assumptions;
for instance, GR depicts gravity as a geometric theory of space-time, which is curved by any form of energy.
For a given distribution of matter, the structure of space-time is determined by the Einstein field equations~\cite{Einstein:1916vd,Einstein:1915ca}.
However, we face a troublesome problem: There is more than one way to describe gravity equivalent to GR. 
GR can be represented in three equivalent representations which are equivalent to each other at their background level, up to a total derivative (boundary term) linear in the Lagrangian: the curvature representation for ordinary GR, which geometrically describes the variation of the orientation of a vector in the parallel transport;
the teleparallel representation for the teleparallel equivalent to general relativity (TEGR), where the location of a vector will change in the parallel transport~\cite{Maluf:2013gaa,Aldrovandi:2013wha};
the nonmetricity representation for the symmetric teleparallel equivalent to general relativity (STEGR), where the length of a vector will change in the parallel transport~\cite{Nester:1998mp,BeltranJimenez:2018vdo,BeltranJimenez:2019esp}.

Moreover, we already know the shortcomings of the above theories; they cannot adequately describe the late-time accelerated expansion of the Universe. 
The situation became more desperate with the improvement of the accuracy of observations~\cite{SupernovaSearchTeam:1998fmf,Boomerang:2000efg,Planck:2018nkj}. 
In order to solve this problem, we have to introduce the concept of dark energy (DE). 
In the framework of the above three basic theories, dark energy manifests itself in the form of exotic matter with negative pressure, implying that it violates the strong energy condition. 
Although the cosmological constant provides the simplest way to account for DE, it also causes notorious problems, such as fine-tuning and coincidence problems.

Another way to avoid such issues is to modify the gravitational theories. 
Straightforward modifications of the Lagrangian of GR, TEGR, and STEGR give us $f(R)$, $f(T)$, and $f(Q)$ gravity, respectively.
These have infinite possibilities due to the arbitrary Lagrangian function, which could provide us with solutions to the DE problem.
This work sheds light on $f(Q)$ gravity~\cite{BeltranJimenez:2017tkd}.
One notable advantage of ordinary GR is that the field equations of $f(Q)$ gravity remain second order of the derivative, regardless of the Lagrangian function. 
Furthermore, it is also notable that $f(Q)$ gravity is not equivalent to $f(R)$ gravity. 
This inequality stems from the total derivative term, which no longer remains linear, and thus the boundary term cannot be dropped by the integral~\cite{BeltranJimenez:2019tme}.
These two properties of $f(Q)$ gravity are the same as those of $f(T)$ gravity,
and one expects different outcomes in $f(Q)$ gravity from those in $f(R)$ or $f(T)$ gravity even for the same function form.

Recently, $f(Q)$ gravity has been intensively investigated. As it applies to cosmology and astrophysics, for instance, cosmological perturbations~\cite{BeltranJimenez:2019tme,Albuquerque:2022eac,Najera:2021afa}, cosmography applications~\cite{Capozziello:2022wgl,Mandal:2020buf}, spherically symmetric solutions~\cite{Wang:2021zaz,DAmbrosio:2021zpm}, and gravitational waves~\cite{Hohmann:2018wxu,Soudi:2018dhv} have been studied. 
For the development of the theory itself, 
the covariant formulation of $f(Q)$ gravity has been investigated in~\cite{Zhao:2021zab,Lin:2021uqa,DAmbrosio:2021zpm}.
Moreover, the construction of a conformal invariant theory~\cite{Gakis:2019rdd} and a generalization to the so-called general teleparallel quadratic tensor theories~\cite{BeltranJimenez:2019odq} and its Hamiltonian analysis have been proposed~\cite{Dambrosio:2020wbi}. To perform a canonical quantization procedure for $f(Q)$ cosmology, a Hamiltonian formulation has been established in the cosmology background (nonperturbation level)~\cite{Dimakis:2021gby}.

Despite extensive research on $f(Q)$ gravity, many aspects of this theory have not yet been studied; for example, it has not been thoroughly investigated how many degrees of freedom (d.o.f.) it possesses. 
This issue is essential because it will not only have a direct effect on its cosmological perturbation theory but it will reveal the implicit symmetry in $f(Q)$ gravity that we might not have noticed before.
The above is the primary purpose of this paper. 

To find out the number of d.o.f. in $f(Q)$ gravity, the best way is to apply the Hamiltonian
formulation~\cite{Henneaux:1992ig,Rothe:2010dzf}. Following Dirac's procedure~\cite{Anderson:1951ta,Dirac:1951zz}, one can analyze the constraint structure strictly and obtain the independent dynamical variables straight from the Lagrange or field equations. 
The above is the approach we take in this paper. 
For the sake of simplicity, we concentrate on the framework of the coincident gauge (CG), in which the inertial connection can be trivialized by diffeomorphism (diff.); hence there is no curvature or inertial concept in coincident general relativity (CGR).
Regarding physical d.o.f., we know that TEGR and CGR have 2 d.o.f due to their equivalence of GR~\cite{Maluf:2000ag,daRochaNeto:2010xls,Ferraro:2016wht,DAmbrosio:2020nqu}. 
On the other hand, the $f(T)$ gravity has 5 d.o.f, and the extra 3 comes from the violation of local Lorentz invariance~\cite{Li:2011rn,Ferraro:2018tpu,Blagojevic:2020dyq}. 
Moreover, in the coincident gauge, the general covariance is broken, which will affect the d.o.f. in general. 
Therefore, physical d.o.f. in $f(Q)$ gravity are highly nontrivial and totally different from those in the $f(R)$ or $f(T)$ gravity theories studied before. 

As was briefly argued, this work potentially contributes to the mode decomposition of the cosmological perturbations as well as gravitational waves in $f(Q)$ gravity.
Although the existing works have shown features of $f(Q)$ gravity with specific models assumed,
the Hamiltonian analysis can clarify the structure of the $f(Q)$ gravity in a model-independent way.
From theoretical viewpoints, the analysis of propagating modes can provide us with a guideline for calculations; that is, how many perturbations we take into account and what kind of perturbations we consider when we go beyond the background-level analysis. 
Moreover, from phenomenological viewpoints, the perturbation or polarization unique to   $f(Q)$ gravity are probes to distinguish $f(Q)$ gravity from other modified gravity theories.
For instance, a comparison of the $f(R)$ and $f(T)$ gravity theories is helpful to reveal the specific feature of $f(Q)$ gravity.
The $f(R)$ and $f(T)$ gravity theories have been investigated for several decades, and we can utilize the known results in these two theories as references for studying $f(Q)$ gravity.

The present paper is organized as follows: The geometric background, the gravitational action, and a brief review of the Hamiltonian formulation of CGR are presented in Sec.~\ref{sec2}.
We investigate Arnowitt-Deser-Misner (ADM) foliation of $f(Q)$ gravity in Sec.~\ref{sec3}.
In Sec.~\ref{sec4} we apply standard Hamiltonian analysis to $f(Q)$ gravity.
In Secs.~\ref{sec5} and Secs.~\ref{sec6}, we count the d.o.f. in $f(Q)$ gravity and give a brief discussion about the physical meaning of our conclusion.
For clarity of notation, we adopt the conventional symbol for the indices in this paper. 
The latin indices $\left(i,j,k, \cdots \right)$ running from 1 to 3 represent the spatial indices, and greek indices $\left( \alpha ,\beta, \cdots \right)$ running from 0 to 3 represent the space-time indices. 
We define some special symbols in Table~\ref{tab:notation}.
\begin{table}[htbp]\renewcommand\arraystretch{1.4}
    \caption{Conventions and notations}
    \label{tab:notation}
    \centering
    \begin{tabular}{|c|l|}
    \hline 
    $\left\{ {}^{\, \alpha}_{\,\mu \nu} \right\}$ & Levi-Civita connection \\
    $\hat\varGamma ^{\alpha}_{\ \mu \nu}$& General affine connection \\
    $\varGamma ^i_{\ jk}$& Levi-Civita connection with respect to $h_{ij}$ \\
    $\nabla _{\mu}$ & Covariant derivative with respect to Levi-Civita connection \\
    $\hat\nabla_{\alpha}$ & Covariant derivative with respect to general affine connection \\
    $\mathcal{D} _i$ & Covariant derivative with respect to $h_{ij}$ \\
    $\mathcal{L}_{\vec{N}}$ & Lie derivative with respect to the vector field \\
    $\hat{R}$ & Curvature scalar with respect to general affine connection \\
    $\mathcal{R}$ & Curvature scalar with respect to Levi-Civita connection \\
    ${}^{3}\mathcal{R}$ & Three-dimensional curvature scalar with respect to Levi-Civita connection \\
    ${}^{3} Q$ & Three-dimensional non-metricity scalar with respect to $h_{ij}$ \\
    ${}^{3}Q_{j k l}$ & Nonmetricy tensor projected onto spatial hypersurface\\
    ${}^{3}Q^{i}$ & Three-dimensional nonmetricy contracted by induced metric, ${}^3{Q^{i}}\equiv {}^3Q_{j k l}h^{i j} h^{k l}$\\
    ${}^{3}\tilde{Q}^{i}$ & Same as ${}^3{Q^{i}}$, ${}^{3}\tilde{Q}^{i} \equiv{}^3\tilde Q_{kjl}h^{ji}h^{k l} $\\
    \hline
    \end{tabular}
\end{table}

\section{Coincident General Relativity} \label{sec2}

In this section, we briefly review the underlying geometrical background that will be used throughout this work in Sec~\ref{subsecA}. 
Then, in order to recall the standard Dirac's procedure for constrained systems in field theory, we also review Hamiltonian formulation and encapsulate the main conclusion of~\cite{DAmbrosio:2020nqu} in~\ref{subsecB}, which has performed a rigorous Hamiltonian analysis of CGR.
%
\subsection{Geometrical foundations}
\label{subsecA}

To start with, we extend the Riemannian geometry by treating the metric and connection as two independent variables. Moreover, the connection is not necessarily symmetric or compatible, which means that besides the curvature, torsion and nonmetricity can also be nonvanishing in the manifold. This implies that the decomposition
of a general affine connection can be written as
\begin{equation}
	\hat\varGamma ^{\alpha}_{\ \mu \nu}=\left\{ {}^{\, \alpha}_{\, \mu \nu} \right\}+K_{\ \mu \nu}^{\alpha}+L_{\  \mu \nu}^{\alpha}~,\label{Eq: Palatini}
\end{equation}
 where $\left\{ {}^{\, \alpha}_{\,\mu \nu} \right\}$ is Levi-Civita connection which can be uniquely determined by the first-order derivatives of the metric
\begin{equation}
	\left\{ {}^{\, \alpha}_{\,\mu \nu} \right\}=\frac{1}{2} g^{\alpha \lambda}\left(g_{\lambda \nu, \mu}+g_{\mu \lambda, \nu}-g_{\mu \nu, \lambda}\right)~.\label{Eq: Levi-Civita}
\end{equation}
Moreover, $K_{\ \mu \nu}^{\alpha}$ and $ L_{\ \mu \nu}^{\alpha} $ are the contortion tensor and deformation tensor respectively, both describing non-Riemannian properties in the manifold
\begin{align}
	K^{\alpha}_{\ \mu \nu} 
	=& \frac{1}{2} g^{\alpha \lambda} \left( T_{\lambda \mu \nu} + T_{\mu \nu \lambda} - T_{\nu \lambda \mu} \right)~, \label{Eq: contorsion}
	\\
    L^{\alpha}_{\ \mu \nu} 
    =& - \frac{1}{2} g^{\alpha \lambda} \left( Q_{\mu \lambda \nu} + Q_{\nu \lambda \mu} - Q_{\lambda \mu \nu} \right)~, \label{Eq: disformation}
\end{align}
where we define the torsion tensor $T_{\ \alpha \beta}^{\mu}$ and the nonmetricity tensor $Q_{\alpha \mu \nu}$ as
\begin{align}
    T_{\ \alpha \beta}^{\mu} &\equiv  \frac{1}{2}(\hat{\varGamma}_{\ \alpha \beta}^{\mu}-\hat{\varGamma}_{\ \beta\alpha}^{\mu})~,\nonumber\\
	Q_{\alpha \mu \nu} &\equiv  \hat\nabla_{\alpha} g_{\mu \nu}=\frac{\partial g_{\mu \nu}}{\partial x^{\alpha}}-g_{\nu \sigma} {\hat\varGamma}_{\mu \alpha}^{\sigma}-g_{\sigma \mu} {\hat\varGamma}_{\nu \alpha}^{\sigma}~.\label{Eq: nonmetricity}
\end{align}
Note that the indexes are raised and lowered by the metric, for example
\begin{equation}
	Q_{\ \mu \nu}^{\alpha}=g^{\alpha \beta} Q_{\beta \mu \nu}=g^{\alpha \beta} \hat\nabla_{\beta} g_{\mu \nu}=\hat\nabla^{\alpha} g_{\mu \nu}~.
\end{equation}

Here and after, we impose the condition that the connection is symmetric, so that the torsion tensor $T_{\ \mu \nu}^{\alpha}$  and the contortion tensor $ K_{\ \mu \nu}^{\alpha}=0$ vanish, and Eq.~\eqref{Eq: Palatini} turns into
\begin{equation}
	\hat\varGamma ^{\alpha}_{\ \mu \nu}=\left\{ {}^{\, \alpha}_{\,\mu \nu} \right\}+L_{\ \mu \nu}^{\alpha}~,\label{Eq: Palatini2}
\end{equation}
so we can rewrite the Riemann curvature tensor in terms of the general connection
\begin{equation}
	\begin{aligned}
	\hat{R}_{\ \beta \mu \nu}^{\alpha}(\hat\varGamma)
&=\partial_{\mu} \hat\varGamma^{\alpha}_{\ \nu \beta}-\partial_{\nu} \hat\varGamma_{\ \mu \beta}^{\alpha}+\hat\varGamma_{\ \mu \lambda}^{\alpha} \hat\varGamma_{\ \nu \beta}^{\lambda}-\hat\varGamma_{\ \nu \lambda}^{\alpha} \hat\varGamma_{\ \mu \beta}^{\lambda}\\
&=\mathcal{R}_{\ \beta \mu \nu}^{\alpha}+ (\partial_{\mu} L_{\ \nu \beta}^{\alpha}- \partial_{\nu} L_{\ \mu \beta}^{\alpha}+L^{\alpha}_{\ \mu \lambda} L_{\ \nu \beta}^{\lambda}-L_{\ \nu \lambda}^{\alpha} L_{\ \mu \beta}^{\lambda})\\
&+\left\{ {}^{\, \alpha}_{\,\mu \lambda} \right\}L_{\ \nu \beta}^{\lambda}+
\left\{ {}^{\,  \lambda}_{\, \nu \beta} \right\}L_{\ \mu \lambda}^{\alpha}
-\left\{ {}^{\,  \lambda}_{\, \mu \beta} \right\}L_{\ \nu \lambda}^{\alpha}
-\left\{ {}^{\, \alpha}_{\,\nu \lambda} \right\}L_{\ \mu \beta}^{\lambda}
+\left(\left\{ {}^{\, \lambda}_{\,\nu\mu} \right\}L_{\ \lambda \beta}^{\alpha}
-\left\{ {}^{\, \lambda}_{\,\mu \nu} \right\}L_{\ \lambda \beta}^{\alpha}\right)\\
&=\mathcal{R}_{\ \beta \mu \nu}^{\alpha}+ (\nabla_{\mu} L_{\ \nu \beta}^{\alpha}- \nabla_{\nu} L_{\ \mu \beta}^{\alpha}+L^{\alpha}_{\ \mu \lambda} L_{\ \nu \beta}^{\lambda}-L_{\ \nu \lambda}^{\alpha} L_{\ \mu \beta}^{\lambda})~,\label{Eq: Riemann tensor1}
\end{aligned}
\end{equation}
where  $\mathcal{R}_{\ \beta \mu \nu}^{\alpha}$ is the usual Riemann tensor in GR composed by the Levi-Civita connection
\begin{equation}
	\mathcal{R}_{\ \beta \mu \nu}^{\alpha}(\left\{ {}^{\, }_{} \right\})=\partial_{\mu} \left\{ {}^{\,\alpha}_{\,\nu \beta} \right\}-\partial_{\nu} \left\{ {}^{\, \alpha}_{\,\mu \beta} \right\}+\left\{ {}^{\, \alpha}_{\,\mu \lambda} \right\} \left\{ {}^{\, \lambda}_{\,\nu \beta} \right\}-\left\{ {}^{\, \alpha}_{\,\nu \lambda} \right\} \left\{ {}^{\, \lambda}_{\,\mu \beta} \right\}~.\label{Eq: Riemann tensor}
\end{equation}
Moreover, the contraction form of~\eqref{Eq: Riemann tensor1} is
\begin{equation}
\hat{R}=\mathcal{R}+(L^{\mu}_{\ \mu \lambda} L_{\ \nu}^{\lambda \ \nu}-L_{\ \nu\lambda }^{\mu} L_{\ \mu}^{\lambda\ \nu})+\nabla_{\mu} L^{\mu\  \nu}_{\ \nu}- \nabla_{\nu} L^{\mu\  \nu}_{\ \mu}~,\label{Eq: relation Q and R}
\end{equation}
noticing that last two terms are total derivative terms which can be eliminated after the integral. 

For the sake of generality, here we will investigate the action of coincident gauge gravity \cite{BeltranJimenez:2017tkd}. We start from action of the general quadratic theory~\footnote{For simplicity, we have ignored the $1/{(16 \pi G)}$ in the front of the action, which will not affect our final results.}~\cite{Heisenberg:2018vsk,BeltranJimenez:2018vdo}
\begin{equation}
\mathcal{S}_{\mathrm{GQ}}
=\int \mathrm{d}^{4} x\left(\sqrt{-g} Q_{\alpha \mu \nu} P^{\alpha \mu \nu} +\lambda_{\alpha}{ }^{\beta \mu \nu} \hat{R}^{\alpha}{ }_{\beta \mu \nu}+\lambda_{a}{ }^{\mu \nu} T^{\alpha}_{\ \mu \nu}\right)~.
\label{Eq: action of the general quadratic theory}
\end{equation}
The last two terms represent the Lagrange multipliers to impose the symmetric teleparallelism condition $\hat{R}^{\alpha}{ }_{\beta \mu \nu}=T^{\alpha}{ }_{\mu \nu}=0$. We define the first term $Q=Q_{\alpha \mu \nu} P^{\alpha \mu \nu}$ as the nonmetricity scalar, where $P^{\alpha \mu \nu}$ is called the superpotential. Defining the trace of the nonmetricity tensor $Q_{\alpha} \equiv Q_{\alpha}{ }_{\mu}{ }^{\mu}$ and  $\tilde{Q}_{\alpha} \equiv Q^{\mu}{}_{\alpha \mu}$, $P^{\alpha \mu \nu}$ is written as 
\begin{align}
	P^{\alpha \mu \nu}
	=c_1Q^{\alpha \mu \nu}+c_2 Q^{\mu \alpha \nu}+c_3Q^{\alpha} g^{\mu \nu}+c_4 g^{\alpha \mu} \tilde{Q}^{\nu}+\frac{c_5}{2}(\tilde{Q}^{\alpha} g^{\mu \nu}+g^{\alpha \nu} Q^{\mu})~.\label{Eq: superpotential}
\end{align}

The symmetric teleparallelism condition restricts the general connection to be total inertia; thus in an arbitrary coordinate system $x^{\alpha}$, the connection should be~\cite{BeltranJimenez:2017tkd}
\begin{equation}
\hat\varGamma ^{\alpha}_{\ \mu \beta}=\frac{\partial x^{\alpha}}{\partial \xi ^{\lambda}}\frac{\partial ^2\xi ^{\lambda}}{\partial x^{\mu}\partial x^{\beta}}~.\label{Eq: inertial connection}
\end{equation}
One can verify this point by substituting~\eqref{Eq: inertial connection}  into~\eqref{Eq: Riemann tensor1}. We can always choose a coordinate $x^{\alpha}=\xi ^{\lambda}$ by utilizing general coordinate transformation, where the general affine connection $\hat\varGamma ^{\alpha}_{\ \mu \nu}=0$. This spacial frame is also known as $coincident \, gauge$. Under such consideration, the Palatini formalism turns into metric formalism, while the metric becomes the only variable. Equation~\eqref{Eq: Palatini2} gives
\begin{align}
L_{\ \mu \nu}^{\alpha} 
= - \left\{ {}^{\, \alpha}_{\,\mu \nu} \right\} 
= -\frac{1}{2} g^{\alpha \lambda}\left(g_{\lambda \nu, \mu}+g_{\mu \lambda, \nu}-g_{\mu \nu, \lambda}\right)~. \label{Eq: disformation tensor2}
\end{align}

As we will show below, under this gauge, in order to meet the requirement that action \eqref{Eq: action of the general quadratic theory} is equivalent to that of GR, one needs to choose the coefficients in the superpotential $P^{\alpha\mu\nu}$ as $c_1=-1/4$, $c_2=1/2$, $c_3=1/4$, $c_4=0$, $c_5=-1/2$, therefore the action \eqref{Eq: action of the general quadratic theory} turns out to be
\begin{equation}
\mathcal{S}_{\mathrm{CGR}}=\frac{1}{4} \int \mathrm{d}^{4} x\sqrt{-g}\left(-Q^{\alpha \nu \rho} Q_{\alpha \nu \rho}+2 Q^{\alpha \nu \rho} Q_{\rho \alpha \nu}-2 Q^{\rho} \tilde{Q}_{\rho}+Q^{\rho} Q_{\rho}\right)~.\label{Eq: action of CGR}
\end{equation}

As is well known, the Hilbert-Einstein action is
\begin{equation}
	\mathcal{S}_{\mathrm{EH}}=\int \mathrm{d}^{4} x \sqrt{-g} \mathcal{R}(\left\{ \right\})~,
	\label{Eq: EH action}
\end{equation}
where $\mathcal{R}(\left\{ \right\})$ is obtained by contracting the Riemann tensor in \eqref{Eq: Riemann tensor}. On the other hand, in standard GR, the second derivatives of the metric in the Ricci scalar can be eliminated by an integration by parts or by adding a suitable boundary or total derivative term, but at the expense of general covariance. 
Dropping the total derivative, one finds the gravitational action~\eqref{Eq: EH action} can be reformulated in a form~\footnote{The first term of action~\eqref{Eq: Hilbert-Einstein action2} is known as Einstein action because Einstein himself formulated it in the first place. Variation of Einstein action toward metric directly gives Einstein tensor without adding any applicable boundary term, for example, Gibbons-Hawking-York boundary term. However, this property is at the expense of general covariance.}
as~\cite{landau1975classical,blau2011lecture}
\begin{equation}
	\mathcal{S}_{{\mathrm{EH}}}=\int d^{4}x \,\sqrt{-g} g^{\mu \nu}\left(\left\{ {}^{\,\alpha}_{\,\sigma \mu} \right\} \left\{ {}^{\,\sigma}_{\,\nu \alpha} \right\}-\left\{ {}^{\,\alpha}_{\,\sigma \alpha} \right\} \left\{ {}^{\,\sigma}_{\,\mu \nu} \right\}\right)+ \int d^{4}x \,\partial _{\alpha}\left( \sqrt{-g} \,\omega ^{\alpha} \right) ,\label{Eq: Hilbert-Einstein action2}
\end{equation}
in which $\omega ^{\alpha}=g^{\mu \nu}\varGamma ^{\alpha}_{\ \mu \nu}-g^{\mu \alpha}\varGamma ^{\nu}_{\ \mu \nu}$. Substituting Eq.~\eqref{Eq: disformation tensor2} with~\eqref{Eq: Hilbert-Einstein action2} and eliminating the boundary term, we will have
\begin{equation}
	\mathcal{S}=-\int \sqrt{-g}g^{\mu \nu}\left(L_{\ \sigma \mu}^{\alpha} L_{\ \nu \alpha}^{\sigma}-L_{\ \sigma \alpha}^{\alpha} L_{\ \mu \nu}^{\sigma}\right) d^{4} x~,\label{Eq: nonmetricity scalar1}
\end{equation}
which is nothing but the nonmetricity scalar in action \eqref{Eq: action of CGR}, namely
\begin{equation}
	Q = g^{\mu \nu}\left(-L_{\ \sigma \mu}^{\alpha} L_{\ \nu \alpha}^{\sigma}+L_{\ \sigma \alpha}^{\alpha} L_{\ \mu \nu}^{\sigma}\right)~.\label{Eq: nonmetricity scalar2}
\end{equation}

But we have to notice, Eq.~\eqref{Eq: nonmetricity scalar1} is not the standard Einstein-Hilbert action as in Eq.~\eqref{Eq: Hilbert-Einstein action2}, and the difference between these two equations exists by the presence of a total derivative term. 
Thus, the nonmetricity scalar and Ricci scalar also differ from each other due to this total derivative term, namely 
\begin{equation}
	\mathcal{R}(\left\{ {}^{\, }_{} \right\})=Q-\nabla_{\alpha}\left(Q^{\alpha}-\tilde{Q}^{\alpha}\right)~.\label{Eq: ricii scalar}
\end{equation}
However, the situation will be significantly different if we move to $f(Q)$ gravity because the total derivative term in the CGR action will no longer be a boundary term in $f(Q)$ gravity.

\subsection{Hamiltonian formulation} \label{subsecB}
In this section, we will encapsulate the Hamiltonian analysis for CGR by following Ref.~\cite{DAmbrosio:2020nqu}. 
In ADM formalism, the space-time metric is written as
\begin{equation}
	d s^{2}=-N^{2} d t^{2}+h_{i j}\left(d x^{i}+N^{i} d t\right)\left(d x^{j}+N^{j} d t\right)~,
\end{equation}
where $ N $, $ N^i $, $ h_{ij} $ are the lapse function, the shift function and the three-dimensional spatial metric. 
Using the above variable, it is straightforward to represent a four-dimensional metric as follows:
\begin{equation}
\begin{aligned}
	g_{\mu \nu} &=\left(\begin{array}{cc}
		-N^{2}+N_{i} N^{i} & N_{i} \\
		N_{i} & h_{i j}
	\end{array}\right)~, \\
	g^{\mu \nu} &=\left(\begin{array}{cc}
		-N^{-2} & N^{-2} N^{i} \\
		N^{-2} N^{i} & h^{i j}-\frac{N^iN^j}{N^2}
	\end{array}\right)~.\label{Eq: ADM metric}
\end{aligned}
\end{equation}
Inserting the above metric into \eqref{Eq: action of CGR} and dropping the boundary terms bring the CGR Lagrangian into the form~\footnote{Partial integration would not affect the numbers of d.o.f of the system, and we shall see in Sec.~\ref{sec3}, the action \eqref{Eq: actionCGR} can be obtained more efficiently by using~\eqref{Eq:  relationship Q3R3} than applying direct decomposition.}
\begin{equation}
\mathcal{S}[g]
=\int \mathrm{d}^{4} x \sqrt{h} N\left[{}^3Q-\mathcal{D}_{l}({}^3{Q^{l}}-{}^3{\tilde{{Q}}^{l}})+K^{i j} K_{i j}-K^{2}\right]~,
\label{Eq: actionCGR}
\end{equation}
where ${}^3Q=h^{ik}\left( \varGamma ^m_{\ il}\varGamma ^l_{\ km}-\varGamma ^l_{\ ik}\varGamma ^m_{\ lm} \right)$ is three-dimensional non-metricity scalar, $K_{i j}=\frac{1}{2 N}\left(\dot{h}_{i j}-\mathcal{D}_{i} N_{j}-\mathcal{D}_{j} N_{i}\right) $ is the extrinsic curvature of the hypersurface, and $K=h^{ij} K_{ij}$ is the trace of $K_{i j}$. However, not all canonical ADM variables in~\eqref{Eq: ADM metric} are independent. They will be related through the primary constraints
\begin{equation}
\begin{aligned}
{\pi}^{N} &=\frac{\delta S}{\delta \dot{N}}=0~, \\
{\pi}^{i} &=\frac{\delta S}{\delta \dot{N}_{i}}=0~.\label{eq: CGR primary constraint}
\end{aligned}
\end{equation}
Therefore, canonical Hamiltonian density can be constructed in the following structure
\begin{equation}
\begin{aligned}
H&=\int_{\varSigma_{t}} \mathrm{~d}^{3} x\left(\lambda {\pi}_{N}+\lambda^{i} {\pi}_{i}+H_{0}\right)\\
&=\int_{\varSigma_{t}} \mathrm{~d}^{3} x\left(\lambda {\pi}_{N}+\lambda^{i} {\pi}_{i}+N {C}_{0}+N^{i} {C}_{i}\right)~,
\end{aligned}
\end{equation}
where $\lambda,\ \lambda^{i}$ are Lagrange multipliers which can be ignored since ${\pi}^{N}$ and ${\pi}^{i}$ are first-class (FC) constraints. 
Moreover, ${C}_{0},\ {C}_{i}$ provide secondary constraints~\cite{DAmbrosio:2020nqu}
\begin{equation}
\begin{aligned}
{C}_{0} &:=-\sqrt{h}\left[{}^3Q-\mathcal{D}_{l}({}^3{Q^{l}-{}^3{\tilde{{Q}}^{l}})}-\frac{1}{h}\left({\pi}_{i j} {\pi}^{i j}-\frac{1}{2}\left({\pi}^{i}{}_{i}\right)^{2}\right)\right]\stackrel{!}{\approx}0~, \\
{C}_{i} &:=-2 \mathcal{D}_{j} {\pi}^{j}{}_{i}\stackrel{!}{\approx}0~.\label{eq: CGR secondary constraint}
\end{aligned}
\end{equation}
We can see that the above has a very similar form to GR. 
The demand for preservation over time requests Poisson brackets (PBs) between the Hamiltonian and secondary constraints equal to zero; this leads to the following equations
\begin{equation}
\begin{aligned}
\left\{{C}_{0}(x), H\right\}=& \int_{\varSigma_{t}} \mathrm{~d}^{3} y\left(N(y)\left\{{C}_{0}(x), {C}_{0}(y)\right\}+N^{i}(y)\left\{{C}_{0}(x), {C}_{i}(y)\right\}\right)~, \\
\left\{{C}_{i}(x), H\right\}=& \int_{\varSigma_{t}} \mathrm{~d}^{3} y\left(N(y)\left\{{C}_{i}(x), {C}_{0}(y)\right\}+N^{j}(y)\left\{{C}_{i}(x), {C}_{j}(y)\right\}\right)~.\label{Eq: consistency conditions}
\end{aligned}
\end{equation}
To obtain the constraint algebra efficiently, we consider the quantities in the following way
\begin{equation}
\begin{aligned}
C_{S}(N) &:=\int_{\varSigma_{t}} N {C}_{0} \mathrm{~d}^{3} x~, \\
C_{V}(\vec{N}) &:=\int_{\varSigma_{t}} N^{i} {C}_{i} \mathrm{~d}^{3} x~.\label{Eq: constraintquantities}
\end{aligned}
\end{equation}
After elaborate calculations, the following PBs can be written in the form of the Lie derivative. After using \eqref{Eq: constraintquantities} and \eqref{eq: CGR secondary constraint}, the constraint algebra turns out to be zero
\begin{equation}
\begin{aligned}
\left\{C_{S}(N), C_{V}(\vec{N})\right\} &=-C_{S}\left(\mathcal{L}_{\vec{N}} N\right)=0~, \\
\left\{C_{V}(\vec{N}_{1}), C_{V}(\vec{N}_{2})\right\} &=C_{V}\left([\vec{N}_{1}, \vec{N}_{1}]\right)=0~, \\
\left\{C_{S}\left(N_{1}\right), C_{S}\left(N_{2}\right)\right\} &=C_{V}\left(\left(N_{1} \partial^{i} N_{2}-N_{2} \partial^{i} N_{1}\right) \partial_{i}\right)=0~.\label{32}
\end{aligned}
\end{equation}

That means the consistency conditions~\eqref{Eq: consistency conditions} are automatically satisfied after using the secondary constraints, and~\eqref{32} show all eight constraints are FC~\footnote{In phase space, a dynamical variable is called FC if it has weakly vanishing PBs with all constraints. If one is not FC, we call it second-class (SC). }, and the Lagrange multipliers $\lambda,\ \lambda^{i}$ remain arbitrary. 
Thus, the d.o.f. of the CGR is $10-8=2$, which is the same as the number of propagating modes with GR. 
This result confirms that the CGR and GR are somewhat equivalent. Actually,~\eqref{Eq: actionCGR} is identical to the Gauss-Codazzi form of the Hilbert-Einstein action, as we will prove in the next section.

\section{ADM decomposition of $f(Q)$ gravity} \label{sec3}
In this section, we attempt to apply the approach developed in the preceding section to $f(Q)$ gravity. 
To establish the $3 + 1$ decomposition of a generic $f(Q)$ gravity theory, we decompose the space-time onto a three-dimensional manifold $\varSigma_{t}$ and express the nonmetricity scalar restricted to the hypersurfaces. 
We start by recalling the corresponding Gauss-Codazzi action in GR. Then, we utilize its results and the relationship between the nonmetricity scalar and Ricci scalar to establish ADM formalism in the $f(Q)$ scenario.\par
In this sense, depending on the corresponding prescription, the direction of time $ T^\mu $ can be decomposed into components tangential and orthogonal to the hypersurface $\varSigma_{t}$
$$T^\mu=N n^\mu+N^\mu~, $$
where $n^\mu$ is defined as a unit vector normal to the hypersurface,  satisfying the normalization condition $n^\mu n_\mu=-1$, and all indices are raised and lowered by $ g_{\mu\nu} $. The explicit form of $n_\mu$ and $n^\mu$ in terms of their components can be written as
\begin{equation}
\begin{aligned}
	n_{\mu} &=\left(\begin{array}{llll}
		-N, & 0, & 0, & 0
	\end{array}\right) ~,\\
	n^{\mu} &=\left(\begin{array}{ll}
		1/N, & -N^{i}/N
	\end{array} \right)~.
\end{aligned}
\end{equation}
We choose the ADM metric Eq.~\eqref{Eq: ADM metric} where the induced metric $ h_{ij} $ lying on $\varSigma_{t}$ orthogonal to normal vector $h^{\mu\nu}n_\nu=0$,
also fulfilling the relation
$$\begin{aligned}
	h_{\mu \nu} &=g_{\mu \nu}+n_{\mu} n_{\nu} =\left(\begin{array}{cc}
		N_{i} N^{i} & N_{i} \\
		N_{i} & h_{i j}
	\end{array}\right)~, \\
	h^{\mu \nu} &=g^{\mu \nu}+n^{\mu} n^{\nu} =\left(\begin{array}{cc}
		0 & 0 \\
		0 & h^{i j}
	\end{array}\right)~,
\end{aligned}$$
hence the four-dimensional Ricci scalar, defined by the connection compatible with the spacetime metric, can be expressed in terms of the intrinsic and extrinsic curvature on the hypersurface $ \varSigma_t $ as follows~\cite{Gomez:2021roj}
\begin{equation}
	\sqrt{-g}\mathcal{R}=\sqrt{h} N\left({ }^{3} \mathcal{R}+K_{i j} K^{i j}-K^{2}\right)+2 \sqrt{-g}\left[\nabla _{\mu}\left(n^{\mu} \nabla _{\nu} n^{\nu}\right)-\nabla _{\nu}\left(n^{\mu} \nabla _{\mu} n^{\nu}\right)\right]~.\label{Eq: ADMRicii}
\end{equation}

In order to facilitate our further calculations, it is convenient to rephrase $ f(Q) $ gravity as a scalar-nonmetricity theory, with the assistance of an auxiliary scalar field~\cite{Jarv:2018bgs}
\begin{equation}
S_{f(Q)}= \int d^{4} x \sqrt{-g}\left[f^{\prime}(\varphi) Q+f(\varphi)-\varphi f^{\prime}(\varphi)\right]~,\label{Eq: scalarnonmetricity}
\end{equation}
where $f$ is an arbitrary function of auxiliary field $\varphi$, and the ${}^\prime$ is the derivative with respect to $\varphi$. On the other hand, we know from Eq.~\eqref{Eq: ricii scalar} that the nonmetricity scalar can be written in terms of the Ricci scalar up to the total derivative term. Inserting~\eqref{Eq: ricii scalar} into~\eqref{Eq: scalarnonmetricity} gives
\begin{equation}
S_{f(Q)}= \int d^{4} x \sqrt{-g}\left[f^{\prime} \mathcal{R}+f^{\prime} \nabla_{\alpha}\left(Q^{\alpha}-\tilde{Q}^{\alpha}\right)+f-\varphi f^{\prime}\right]~.
\end{equation}
Combining it with~\eqref{Eq: ADMRicii}, we can express the $ f(Q) $ action by the spatial curvature scalar in following form
\begin{equation}
\begin{aligned}
S_{f(Q)}=& \int d^{4} x \sqrt{-g}\left[f^{\prime}\left({ }^{3} \mathcal{R}+K_{i j} K^{i j}-K^{2}+2 \nabla_{\mu}\left(n^{\mu} \nabla_{\nu} n^{\nu}\right)-2 \nabla_{\nu}\left(n^{\mu} \nabla_{\mu} n^{\nu}\right)\right)\right.\\
&\left.+f^{\prime} \nabla_{\alpha}\left(Q^{\alpha}-\tilde{Q}^{\alpha}\right)+f-\varphi f^{\prime}\right]~.\label{Eq: fQaction}
\end{aligned}
\end{equation}

Furthermore, we prove that there exists an analogous relationship between the intrinsic nonmetricity scalar ${}^3Q$ and the intrinsic curvature ${}^3\mathcal{R}$, which we have explicitly shown in Appendix~\ref{appendix A}
\begin{equation}
{}^3\mathcal{R}={}^3Q-\mathcal{D} _l({}^3{Q^{l}}-{}^3{\tilde{{Q}}^{l}})~.\label{Eq:  relationship Q3R3}
\end{equation}
Inserting~\eqref{Eq:  relationship Q3R3} into~\eqref{Eq: fQaction} and after tedious calculations to the boundary terms presented in Appendix~\ref{appendix B}, we give the action to the following form 
\begin{equation}
	\begin{aligned}
		S_{f(Q)}&= \int d^{4} x\left\{  N\sqrt{h}\left[f+ f'\left({}^3Q+K_{i j} K^{i j}-K^{2}-\varphi\right)-\mathcal{D}_{l}[f'({}^3{Q^{l}}-{}^3{\tilde{{Q}}^{l}})]\right]\right.\\
		&\quad\left.+\frac{\sqrt{h}}{N}\dot{\varphi}f''\left( \partial _iN^i \right) -\frac{\sqrt{h}}{N}\dot{N}^i\partial _if'-\frac{\sqrt{h}}{N}\partial _if'\left( N^i\partial _jN^j-N^k\partial _kN^i \right)\right\}~.\label{Eq: fQ ADM action}
	\end{aligned}
\end{equation}
This is the final decomposed action of $f(Q)$ gravity which we will pursue in the following sections.

\section{Hamitonian formulation in $f(Q)$ gravity} \label{sec4}

In this section, we apply the standard Dirac-Bergmann algorithm~\cite{Dirac:1951zz,Anderson:1951ta} to $f(Q)$ gravity. First of all, it is useful to note that the relation of $\dot{N}^i$ and $\dot{N_i}$ can be expressed as 
\begin{equation}
	\begin{aligned}
\dot{N}^i\partial _if'=\partial _if'\dot{(N_jh^{ij})}&=\partial _if'\dot{N_j}h^{ij}+\partial _if'N_j\dot{h}^{ij}\\
&=\partial ^if'\dot{N_i}-\partial ^if'N^j\dot{h}_{ij}~. \label{eq. relation}
\end{aligned}
\end{equation}
Moreover, by defining $A^{ij}:=\frac{\sqrt{h}}{N}N^j\partial ^if'$ and $   B^{i}:=\left( N^i\partial _jN^j-N^j\partial _jN^i \right) $, we rephrase action~\eqref{Eq: fQ ADM action} into following form
\begin{equation}
	\begin{aligned}
		S_{f(Q)}&= \int d^{4} x \left\{  N\sqrt{h}\left[f+ f'\left({}^3Q+K_{i j} K^{i j}-K^{2}-\varphi\right)-\mathcal{D}_{l}[f'({}^3{Q^{l}}-{}^3{\tilde{{Q}}^{l}})]\right]\right.\\
		&\quad \left.+\frac{\sqrt{h}}{N}\dot{\varphi}f''\left( \partial _iN^i \right) -\frac{\sqrt{h}}{N} \left(\partial ^if'\right)\dot{N_i}+A^{ij}\dot{h}_{ij}-\frac{\sqrt{h}}{N} \left(\partial _if'\right) B^{i}\right\}~.
	\end{aligned}
	\label{Eq: fQ ADM action2}
\end{equation}
There are 11 dynamical variables in total, $ N $, $  N_i $, $ h_{ij} $ , $  \varphi $. Note that $N_i$ has three components, and $h_{ij}$ has six components because of the index symmetry. In order to perform the Hamiltonian formalism of a dynamical system, we introduce, in the usual way, the corresponding conjugate momenta which are defined by
\begin{equation}
	\begin{aligned}
		\pi^{N} &\equiv\frac{\delta S}{\delta \dot{N}}=0~, \\
		\pi^{i} &\equiv\frac{\delta S}{\delta \dot{N}_{i}}=-\frac{\sqrt{h}}{N}\partial ^if' ~,\\
			\pi^{ij} &\equiv\frac{\delta S}{\delta \dot{h}_{i j}}=\sqrt{h}\left[f^{\prime}\left(K^{i j}-h^{i j}K\right)\right]+A^{ij}~,\\
		p &\equiv\frac{\delta S}{\delta \dot{\varphi}}=\frac{\sqrt{h}}{N}f''\left( \partial _iN^i \right)~.\label{Eq: momenta}
	\end{aligned}
\end{equation}
The fundamental PBs~\footnote{The PBs between two functions $F(x)$ and $G(x)$ of the phase-space variables $\left\{N, N^{i}, h_{i j},\varphi, {\pi}_{N}, {\pi}_{i}, {\pi}^{i j},p\right\}$ is defined as \cite{DAmbrosio:2020nqu}: $$\{F(x), G(y)\}:=\int_{\varSigma_{t}} \mathrm{~d}^{3} z \sum_{k}\left(\frac{\delta F(x)}{\delta \Phi^{k}(z)} \frac{\delta G(y)}{\delta {\Pi}_{k}(z)}-\frac{\delta G(y)}{\delta \Phi^{k}(z)} \frac{\delta F(x)}{\delta {\Pi}_{k}(z)}\right)\label{Eq: Possion Bracket}$$} among the canonical variables in phase space are given by
\begin{equation}
\begin{aligned}
\left\{N(x), {\pi}_{N}(y)\right\} &=\delta^{(3)}(\vec{x}-\vec{y})~, \\
\left\{N^{i}(x), {\pi}_{j}(y)\right\} &=\delta_{j}^{i} \delta^{(3)}(\vec{x}-\vec{y})~, \\
\left\{h_{i j}(x), {\pi}^{m n}(y)\right\} &=\delta_{(i}^{m} \delta_{j)}^{n} \delta^{(3)}(\vec{x}-\vec{y})~.
\end{aligned}
\end{equation}

According to the Dirac-Bergmann algorithm, the conjugate momenta in Eq.~\eqref{Eq: momenta} will become constraints of the system if they do not contain the corresponding velocities, which means that the latter cannot be obtained inversely. To see how many constraints there are in Eq.~\eqref{Eq: momenta}, we check the following matrix~\footnote{Some papers call it Hessian: $\mathcal{H}_{ab}=\frac{\partial^2 L}{\partial \dot{q}^{a}\partial \dot{q}^{b}}$, which is exactly same as matrix: $W_{a b}$.}
\begin{equation}
	W_{a b}=\frac{\partial \Pi_{a}}{\partial \dot{q}^{b}}=
	 \begin{array}{c@{\hspace{-5pt}}l}
	\left(
	\begin{array}{c|ccc|ccc|c}
		0 & 0 & \cdots & 0 & 0 & \cdots & 0 & 0 \\
		\hline
		0 & 0 & \cdots & 0 & 0 & \cdots & 0 & 0 \\
		\vdots & \vdots & \ddots & \vdots & \vdots & \ddots & \vdots & \vdots \\
		0 & 0 & \cdots & 0 & 0 & \cdots & 0 & 0 \\
		\hline
		0 & 0 & \cdots & 0 & f_{1} & \cdots & 0 & 0 \\
		\vdots & \vdots & \ddots & \vdots & \vdots & \ddots & \vdots & \vdots \\
		0 & 0 & \cdots & 0 & 0 & \cdots & f_{6} & 0 \\
		\hline
		0 & 0 & \cdots & 0 & 0 & \cdots & 0 & 0
	\end{array}
	\right)
	&\begin{array}{l}\left. \rule{0mm}{7mm}\right\}{ \text{\scriptsize $3$} }\\
	\\ \left. \rule{0mm}{7mm}\right\}{ \text{\scriptsize $6$} } \vspace{0mm}
	\end{array}\\[-5pt]
	\begin{array}{cc}\underbrace{\rule{12mm}{0mm}}_{3} \ & \ \underbrace{\rule{13mm}{0mm}}_{6}\hspace{0pt}
	\end{array}&
	\end{array}, \qquad
	\begin{array}{cccccccc}
		a= (\overbrace{0 }^{\pi ^{N}},  \overbrace{0, \ldots, 0}^{\pi ^{i}} ,\overbrace{0, \ldots, 0}^{\pi ^{ij}} , \overbrace{0}^{p})
	   \end{array}
	\end{equation}
	
Note that  $ W_{ab} $ is a $ 11\times 11 $ symmetric matrix with all components zero, except for those six diagonal elements. So the rank of $ W_{ab} $ is $M=6$, that give rise to $ 11-6=5 $ primary constraints
    \begin{equation}
    	\begin{aligned}
    		\phi _{a}&:\pi _{N}\stackrel{!}{\approx}0~,	\\
    		\phi _{i}&:\pi_{i} +\frac{\sqrt{h}}{N}\partial _if'\stackrel{!}{\approx}0~,\\
    		\phi _{c}&:p-\frac{\sqrt{h}}{N}f''\left( \partial _iN^i \right)\stackrel{!}{\approx}0~,\label{Eq.primary constrains}
    	\end{aligned}		
    \end{equation}
where the symbol $!$ stands for demand. The linear combination of the primary constraints~\eqref{Eq.primary constrains} defining the  hypersurface intersection $\varSigma _1$, guarantees the phase-space variables are weakly equal~\footnote{If the restriction of phase-space variables vanishes on $\phi \left( p,q \right) $, we say $\phi \left( p,q \right) $ is weakly equal to zero, and we label it by $\approx $. In addition, if function $\phi \left( p,q \right) $ and all its first derivatives vanish on $\varSigma _1$, then we say $\phi \left( p,q \right) $ is strongly equal to zero, labeled by $=$.} to zero. 

The independence checking~\footnote{Under general consideration, the primary constraints~\eqref{Eq.primary constrains} may be reducible; thus we can split them into independent part and an dependent part. If the matrix $J_{n'n}$ is full rank then it is called as a regularity condition.} of these primary constraints are given by Jacobian matrix $J_{n'n}=\frac{\partial \phi _{n'}}{\partial \left( q^n,p_n \right)}$. If the rank of matrix $J_{n'n}$ equals $M'$, it means there only exist $M'$ independent primary constraints among~\eqref{Eq.primary constrains}~\cite{Henneaux:1992ig}. In our case, it is not hard to check $J_{n'n}$ is full rank matrix, so that  $M'=5$, it suggests the $\varSigma _1$ is a phase-space submanifold of dimension $2N-M'=17$.

Since $\phi _{i}$ and $\phi _{c}$ are dependent on $N$ and $\varphi$, all PBs between different primary constraints remain nonvanishing. The nonvanishing PBs are listed below
\begin{equation}
	\begin{aligned}
\left\{\phi_{a}(x), \phi_{i}(y)\right\}&= \frac{\sqrt{h}}{N^2}\partial _if' \,\delta^{3}(x-y)~,\\
\left\{\phi_{a}(x), \phi_{c}(y)\right\}&=-\frac{\sqrt{h}}{N^2}f''\left( \partial _iN^i \right) \delta^{3}(x-y)~,\\
\left\{\phi_{c}(x), \phi_{i}(y)\right\}&=-\frac{\sqrt{h}}{N} f^{\prime \prime \prime} \partial_{i} \varphi\;\delta^{3}(x-y)-2 \frac{\sqrt{h}}{N} f^{\prime \prime}\partial_{i}\delta^{3}(x-y) ~.
	\end{aligned}
	\label{Eq: PB1}
\end{equation}

On the other hand, the remaining six equations can be inversely solved to get $ \dot{h_{ij}} $ . Defining $\pi =h_{ij}\pi ^{ij}$ and $A=h_{ij}A^{ij}$, from~\eqref{Eq: momenta} we can get: $\pi =-2\sqrt{h}f'K+A$,  $ K=\frac{A-\pi}{2\sqrt{h}f'} $. Substituting back to~\eqref{Eq: momenta} yields the following relationships
\begin{equation}
\begin{aligned}
  K^{ij}&=\frac{1}{\sqrt{h}f'}\left[ \pi ^{ij}-A^{ij}+\frac{1}{2}h^{ij}\left( A-\pi  \right) \right]~,\\
  K_{ij}K^{ij}-K^2&=\frac{1}{hf'^2}\left[ \left( \pi ^{ij}-A^{ij} \right) \left( \pi _{ij}-A_{ij} \right) -\frac{1}{2}\left( A-\pi \right) ^2 \right]~,
\end{aligned}	
\end{equation} 
hence 
\begin{equation}
  \begin{aligned}
	\dot{h}_{i j}&=2 N K_{i j}+2 \mathcal{D}_{(i} N_{j)}\\
  &=  \frac{2N}{\sqrt{h}f'}\left[\pi_{i j}-A_{ij}+\frac{1}{2} h_{i j}(A-\pi)\right]+ 2\mathcal{D}_{(i} N_{j)}~.
\end{aligned}
\end{equation}

Now we return to further development of the Hamiltonian formalism. We define the total Hamiltonian in the traditional form
\begin{equation}
\begin{aligned}
	H&=\int_{\varSigma_{t}} \mathscr{H} \,d^{3} x \\
	&=\int_{\varSigma_{t}} d^{3} x \left(\lambda ^{a}\pi _{N}+\lambda^{i} \phi_{i}+\lambda ^{c} \phi_{c}+ {\mathscr{H}_0}\right)~,
\end{aligned}
\end{equation}
in which ${\mathscr{H}_0}$ represents the canonical Hamiltonian density
\begin{align}
	 {\mathscr{H}_0}:&=\pi^{ij}\dot{h}_{i j}+\pi^i\dot{N}_{i}+p\dot{\varphi}-\mathscr{L}\nonumber \\
	&=N {C}+2 \mathcal{D}_{(i} N_{j)} \left(\pi^{i j}-A^{ij}\right)+\frac{\sqrt{h}}{N} \partial_{i}       f^{\prime} B^{i}\nonumber \\
    &=N {C}+2 \mathcal{D}_{i} N_{j} \pi^{i j}
      -\frac{\sqrt{h}}{N}\partial _if'\left(N^j\partial ^iN_j-N^i\partial _jN^j+2N^j\partial _jN^i\right)~,\label{Eq:Hamiltonian}
      \nonumber\\
    C&=\frac{1}{\sqrt{h}f'}\left[ \left( \pi ^{ij}-A^{ij} \right) \left( \pi _{ij}-A_{ij} \right) -\frac{1}{2}\left( A-\pi \right) ^2 \right]\nonumber\\
 &\quad-\sqrt{h}\{f+f'\left({}^3Q-\varphi\right)-\mathcal{D}_{l}[f'({}^3{Q^{l}}-{}^3{\tilde{{Q}}^{l}})] \}~,
\end{align}
while $\lambda ^{a}$, $\lambda ^{i}$ and $\lambda ^{c}$ are arbitrary Lagrangian multipliers. In order to find out the evolution of the constraints \eqref{Eq: momenta}, we need to check whether the consistency conditions for these constraints could be satisfied. 
The consistency conditions require that the time derivative of these constraints, which can be transformed into the PBs between the constraint functions with the Hamiltonian, also vanish on a new hypersurface $\varSigma_2$ of lower dimension than $\varSigma_1$. This leads to the following equations:
\begin{equation}
	\begin{aligned}
		\dot{\phi} _{n'}&=\left\{\phi_{n'}, \mathscr{H}\right\}\\
		&\approx\left\{\phi_{n'},  {\mathscr{H}_0}\right\}+\left\{\phi_{n'}, \phi_{n}\right\} \lambda^{n}\stackrel{!}{\approx} 0~, ~n\in \left( a,i,c \right)~.
 \end{aligned}
 \end{equation}
Such conditions can be satisfied if they have solutions for all the multipliers. To check this, we have to compute the rank of the following matrix:
\begin{equation}
C_{n'n}=\left\{ \phi _{n'},\phi _n \right\}  =\left( \begin{matrix}
	0&		A_{1}&		A_2&	A_3&  -B \\
	-A_1&		0&		0&		0&    C_1 \\
	-A_2&		0&		0&		0&      C_2\\
	-A_3&		0&		0&		0&     C_3\\
	B&		-C_1&		-C_2&		-C_3&     0
\end{matrix} \right)~,
\end{equation}
where we define
\begin{equation}
\frac{\sqrt{h}}{N}f''\left( \partial _iN^i \right):=S~,~\frac{\sqrt{h}}{N}\partial _if':=V_i~,~
\left\{\pi _{N},V_i\right\}\equiv A_i=(A_1, A_2 , A_3)~,~\left\{V_{i}, p\right\}+\left\{S,\pi_{i}\right\}\equiv C_i=(C_1, C_2 , C_3)~.
\end{equation}

We find the $ \det(C_{n'n})=0 $ and its rank is 4,  which is consistent with the fact that $ C_{n'n} $ is an antisymmetric matrix of odd order. This means the multipliers are not yet uniquely determined, and we need to find another constraint condition. Such a constraint condition can be made by contracting the null eigenvector of matrix $ C_{n'n} $ 
\begin{equation}
	\xi ^{n}=\left( \begin{matrix}0, &A_{[3}C_{2]},& A_{[1}C_{3]},& A_{[2}C_{1]},& 0\end{matrix} \right) ~,
\end{equation}
and $ h_{n}\equiv (\left\{\phi_{n},  {\mathscr{H}_0}\right\})$, namely, 
\begin{equation}
	\begin{aligned}
		\chi=\xi ^{n} h_{n}=\xi ^{i}\left(\left\{\pi_{i} , \mathscr{H}_0\right\}+\left\{V_{i} , \mathscr{H}_0\right\}\right)
		\stackrel{!}{\approx}0~.\label{eq, secondary constraint}
	\end{aligned}
\end{equation}
The extra constraint imposed by the null eigenvector should be the secondary constraint \cite{Can-BinLiang:2000,blagojevic2001gravitation}.

The presence of secondary constraints are independent of previous constraints $\phi_{n} $, and restrict the motion in phase space to a new hypersurface $\varSigma_2$ of lower dimension than $\varSigma_1$. Now, we have five primary constraints and one secondary constraint, these six constraints ought to be preserved during the evolution of the system. Hence the consistency equation for the above six constraints leads to
\begin{align}
	0 &\approx 
	\dot{\phi}_{a}
	\nonumber \\
	&=\left\{\phi_{a}, \mathscr{H}_0\right\}+\left\{\phi_{a}, \phi_{n}\right\} \lambda^{n} 
	\nonumber \\
	&=\left\{\pi_{N}, \mathscr{H}_0\right\}+\left\{\pi_{N}, \phi_{i}\right\} \lambda^{i}+\left\{\pi_{N}, \phi_{c}\right\} \lambda^{c} \label{Eq: CE1}
	\nonumber \\
	&=\left\{\pi _{N}, \mathscr{H}_0\right\}+\left\{\pi _{N}, V_{i}\right\} \lambda^{i}-\left\{\pi _{N},S\right\} \lambda^{c} ~,
	\\
	0 &\approx 
    \dot{\phi}_{i}
    \nonumber \\
    &=\left\{\phi_{i}, \mathscr{H}_0\right\}+\left\{\phi_{i}, \phi_{n}\right\} \lambda^{n}
    \nonumber \\
	& = \left\{\pi_{i}, \mathscr{H}_0\right\}+ \left\{V_{i} , \mathscr{H}_0\right\}-\left\{ \pi _{N},V_{i}\right\} \lambda^{a}+\left\{S ,\pi _{i} \right\} \lambda^{c}+\left\{V_{i}, p\right\} \lambda^{c} ~,\label{Eq: CE2}
	\\
    0 &\approx 
    \dot{\phi}_{c}
    \nonumber \\
    &=\left\{\phi_{c}, \mathscr{H}_0\right\}+\left\{\phi_{c}, \phi_{n}\right\} \lambda^{n}
    \nonumber \\
    &=\left\{p, \mathscr{H}_0\right\}- \left\{S , \mathscr{H}_0\right\}
		+\left\{\pi _{N}, S\right\} \lambda^{a}-\left\{S , \pi_{i}\right\} \lambda^{i}
		-\left\{V_{i}, p\right\} \lambda^{i} ~,\label{Eq: CE3}
	\\
	 0 &\approx 
	\dot{\chi}
	\nonumber \\
	&=\left\{\chi, \mathscr{H}_0\right\}+\left\{\chi, \phi_{n}\right\} \lambda^{n} ~.\label{Eq: new consistency}
\end{align}

From the above consistency evolution treatment of constrained dynamics, we hope that some initially undermined Lagrange multipliers can be eventually determined by Eqs.~\eqref{Eq: CE1}-\eqref{Eq: new consistency}. 
We will check this point in the next section.

\section{Degrees of freedom of $f(Q)$ gravity} \label{sec5}

\subsection{Counting the degrees of freedom} 

In this section, we will test whether the hypersurface surfaces $\overline{\varSigma }=\varSigma_{1}\cap \varSigma_{2}$~\footnote{$\overline{\varSigma }$ is the constraint surface determined by Eqs.~\eqref{Eq.primary constrains} and the Eq.~\eqref{eq, secondary constraint} .}are final constraint surfaces. This consideration drive us to combine the consistency condition of secondary constraint~\eqref{Eq: new consistency} with the previous $ C_{n'n} $ to consist of a new matrix $ \Phi_{m n} $ 
\begin{equation}
	\left(\Phi_{m n}\right) \equiv\begin{bmatrix}
		C_{n'n} \\
		\left\{\chi, \phi_{n}\right\}
	\end{bmatrix}=\left( \begin{matrix}
	0&		A_{1}&		A_2&	A_3&  -B \\
	-A_1&		0&		0&		0&    C_1 \\
	-A_2&		0&		0&		0&      C_2\\
	-A_3&		0&		0&		0&     C_3\\
	B&		-C_1&		-C_2&		-C_3&     0\\
	\left\{\chi, \phi_{a}\right\}&\left\{\chi, \phi_{x}\right\}&\left\{\chi, \phi_{y}\right\}&\left\{\chi, \phi_{z}\right\}&\left\{\chi, \phi_{c}\right\}\\
\end{matrix} \right) ~.
\end{equation}

At this stage, we are facing two possibilities. If the rank of $ \Phi_{m n} $  is still 4, it means there must exist a column vector $\zeta^n$ which satisfying the relation $C_{n'n}\zeta^n\approx0 $ and $\left\{\chi, \phi_{n}\right\}\zeta^n\approx0 $, leads us to construct a new constraint $\phi_{0}=\zeta^n\phi_{n}$. By a linear combination of primary constraints, one can prove this constraint $\phi_{0}$ must be FC~\cite{Can-BinLiang:2000,Henneaux:1992ig}. 
Then, the consistency condition of $\phi_{0}$ must be checked further.

On the other hand, if the rank of $ \Phi_{m n} $  is maximum, i.e., equal to 5, 
it means the consistency conditions~\eqref{Eq: CE1}-\eqref{Eq: new consistency} contain five linearly independent equations for the Lagrange multipliers, so that the five Lagrange multipliers can be solved as a function of the phase-space variables. 
Inserting these Lagrange multipliers back into Eqs.~\eqref{Eq: CE1}-\eqref{Eq: new consistency}, 
we will certainly see that the consistency conditions for all constraints are automatically satisfied. Based on the above discussion, we check the specific form of Eq.~\eqref{eq, secondary constraint}
\begin{equation}
	\begin{aligned}
		\chi&=\xi ^{i}\left(\left\{\pi_{i} , \mathscr{H}_0\right\}+\left\{V_{i} , \mathscr{H}_0\right\}\right)\\
		&=\xi ^{x}\left(\left\{\pi_{x} , \mathscr{H}_0\right\}+\left\{V_{x} , \mathscr{H}_0\right\}\right)+\xi ^{y}\left(\left\{\pi_{y} , \mathscr{H}_0\right\}+\left\{V_{y} , \mathscr{H}_0\right\}\right)+\xi ^{z}\left(\left\{\pi_{z} , \mathscr{H}_0\right\}+\left\{V_{z} , \mathscr{H}_0\right\}\right)\\
		&=(A_3C_2-A_2C_3)\left\{\pi_{x}+V_{x} , \mathscr{H}_0\right\}+(A_1C_3-A_3C_1)\left\{\pi_{y}+V_{y} , \mathscr{H}_0\right\}+(A_2C_1-A_1C_2)\left\{\pi_{z}+V_{z} , \mathscr{H}_0\right\}~.\label{Eq,expand chi}
	\end{aligned}
\end{equation}

Although very complicated in the details, if we analyze~\eqref{Eq,expand chi} carefully we will find that it possesses good symmetry in its components. The possibility that $ \left(\Phi_{m n}\right) $ becomes a reducible (nonfull rank) case requires at least the $ \left\{\chi, \phi_{x}\right\},\left\{\chi, \phi_{y}\right\},\left\{\chi, \phi_{z}\right\} $ vanish simultaneously, which we note are very strict conditions. For example, $ \phi_{x}=\pi_{x}+ \frac{\sqrt{h}}{N}\partial _xf' $ means  that  $ \chi $ cannot contain the corresponding $ N^{x}, \pi_{ij} $, and $\pi_{N}$ . This is very easily disproved: We consider the first term of $\chi$ from~\eqref{Eq:Hamiltonian}, variate $\mathscr{H}_{0}$ with respect to $N^x$ and $\pi_{ij}$ respectively
\begin{equation}
  \begin{aligned}
\int_{\varSigma_{t}}\left.\delta \mathscr{H}_{0}\right|_{N^x}&=-\int_{\varSigma_{t}}2\sqrt{h}\mathcal{D}_j\left( \frac{\pi ^{j}_{\ x}}{\sqrt{h}} \right)\delta{N^x}-\int_{\varSigma_{t}}F_{x}(h_{ij},N,\varphi,N^{i})\delta{N^x}~,\\
\int_{\varSigma_{t}}\left.\delta \mathscr{H}_{0}\right|_{\pi_{ij}}&=\int_{\varSigma_{t}} \frac{N}{\sqrt{h}f'}\left( 2\pi ^{ij}-\pi h^{ij} \right) \delta \pi _{ij}+\int_{\varSigma_{t}} 2\mathcal{D}^iN^j\delta \pi _{ij}~,
 \end{aligned}
 \end{equation}
where $F_{x}$ is a function of $h_{ij}$, $N$, $\varphi$ and $N^{i}$ with a very complicated form. Then we can express the following PBs as
\begin{equation}
	\begin{aligned}
\left\{\pi_{x}(x), \mathscr{H}_0(y)\right\}&=\left[ 2\sqrt{h}\mathcal{D}_j\left( \frac{\pi ^{j}_{\ x}}{\sqrt{h}} \right)+F_{x}(h_{ij},N,\varphi,N^{i})\right]\delta^{3}(x-y)~,\\
  \left\{V_{x}(x) , \mathscr{H}_0(y)\right\}&= \left( \frac{\pi}{2}-\sqrt{h}\mathcal{D}_jN^j \right) \partial _x\ln f' \, \delta^{3}(x-y)~.
	\end{aligned}
\end{equation}  
We can see that $ \left\{\pi_{x}+V_{x} , \mathscr{H}_0\right\} $ contains terms like $ N^{j}, \pi $ and so on, while the term $(A_3C_2-A_2C_3)$ is nonvanishing. This indicates that $ N^{x}, \pi_{ij} $ must appear in $\chi$, which gives rise to
\begin{equation}
    \left\{\chi , \phi _{i}\right\} \ne 0~.
    \label{Eq: PB2}
\end{equation}
Therefore, the rank of $ \Phi_{m n} $ is 5, which shows that all five Lagrange multipliers can be uniquely determined by~\eqref{Eq: CE1}-\eqref{Eq: new consistency}, thus there is no further secondary constraint, all consistency conditions are exhausted, and the algorithm is finished. 
Moreover, by Eqs. \eqref{Eq: PB1} and \eqref{Eq: PB2}, we know by definition that all the primary and secondary constraints are SC. Finally, we can determine the number of d.o.f. in $f(Q)$ gravity,

\begin{equation}
	\begin{aligned}
		\bm{\mathcal{D}.o.f.}  &= \frac{1}{2}\cdot\left(\begin{array}{c}
\text { Number of original} \\
\text { canonical variables}
\end{array} 
- 
\begin{array}{c}
\text {Total number } \\
\text {of constraints }
\end{array}- 
\begin{array}{c}
\text {Number of } \\
\text {gauge conditions }
\end{array} \right) \\
&= \frac{1}{2}\cdot\left(\begin{array}{c}
\text { Number of original} \\
\text { canonical variables}
\end{array} 
- 2 \times 
\begin{array}{c}
\text {Number of} \\
\text {FC constraints}
\end{array}- 
\begin{array}{c}
\text {Number of } \\
\text {SC constraints }
\end{array} \right)\\
&=\frac{1}{2}\cdot (22-0-6)=8~.
	\end{aligned}
\end{equation}

\subsection{Interpretation of degrees of freedom} 

In this subsection, we try discuss about the origin of 8 d.o.f. in $f(Q)$ gravity. 
and then, we make a comparison between $f(Q)$ with the other two in the geometrical trinity of gravity named $f(T)$ and $f(R)$ gravity. 
Starting from Sec. \ref{sec2}, we have shown that there are eight FC constraints in CGR, which cause all the Lagrange multipliers to stay arbitrary. 
As a consequence, the evolution of dynamical variables in phase space cannot be uniquely determined by their initial values, and it corresponds to the fact that CGR possesses the same gauge symmetry as that in GR. 
As a matter of fact, four constraints~\eqref{eq: CGR secondary constraint} describe general coordinate transformation, and the rest og the four FC constraints~\eqref{eq: CGR primary constraint} can be utilized to fix the nondynamical fields $N$ and $N^{i}$. 
Then it reduces the total variables from ten to two.

However, when we extend the action of CGR to the general case, the situation becomes quite different: When we remove the inertial connection $  \hat\varGamma_{\ \mu \beta}^{\alpha}=\frac{\partial x^{\alpha}}{\partial \xi^{\rho}} \partial_{\mu} \partial_{\beta} \xi^{\rho}$ by means of diff. ($x^{\alpha}=\xi^{\rho}$), the connection vanishes, and the nonmetricity $Q_{\alpha \mu \nu}$ is no longer a tensor in general sense (it does not follow the transformation rules under the general coordinate transformation), and the action is only invariant when the boundary term shows up~\cite{Zhao:2021zab}. 
We can see this point more clearly if we consider the arbitrary gauge $\hat\varGamma_{\ \mu \beta}^{\alpha}=\frac{\partial x^{\alpha}}{\partial \xi^{\rho}} \partial_{\mu} \partial_{\beta} \xi^{\rho} $ and coincident gauge $\hat\varGamma_{\mu \beta}^{\alpha}=0$ for two circumstances. 
By using~\eqref{Eq: Riemann tensor1}, each condition gives us noncurvature condition. Combining Eqs.~\eqref{Eq: nonmetricity scalar2} and~\eqref{Eq: relation Q and R} yields
  \begin{equation}
    \mathcal{R}+\nabla_{\mu}\left(\tilde{Q}^{\mu}-{
        {Q}}^{\mu}\right)-{Q}=\mathcal{R}+\nabla_{\mu}\left(\tilde{\hat{\mathcal{Q}}}^{\mu}-\hat{\mathcal{Q}}^{\mu}\right)-\mathcal{\hat{Q}}=0~.
    \end{equation}
To differentiate, here we use $\mathcal{\hat{Q}}(g_{\mu\nu}, \hat\varGamma_{\ \mu \beta}^{\alpha})$ to represent the nonmetricity scalar for the arbitrary gauge and $Q(g_{\mu\nu},0)$ for coincident gauge. It follows that
\begin{equation}
\sqrt{-g}\,Q(g_{\mu\nu},0)=\sqrt{-g}\,\hat{\mathcal{Q}}(g_{\mu\nu}, \hat\varGamma_{\ \mu \beta}^{\alpha})+\partial_{\mu}\left[\sqrt{-g}({\mathcal{B}}^{\mu}-\hat{\mathcal{B}}^{\mu})\right]~,
  \end{equation}
in which $\mathcal{B}^{\mu}(g_{\mu\nu},0)=\tilde{{Q}}^{\mu}-{{Q}}^{\mu}$ and   $\hat{\mathcal{B}}^{\mu}(g_{\mu\nu},\hat\varGamma_{\ \mu\beta}^{\alpha})=\tilde{\hat{\mathcal{Q}}}^{\mu}-\hat{\mathcal{Q}}^{\mu}$. It explicitly shows that the boundary term eliminated by integration and gives a covariant $ Q(g_{\mu\nu},0) $, but this is only for the case when $f_{QQ}=0 $.

Let us consider the general case of $f(Q)$ gravity, where namely $f_{QQ}$ is not necessarily zero. 
In this case, the action is no longer diff. invariant
$x'^{\mu}=x^{\mu}+\varepsilon ^{\mu}\left( x \right)$ in the coincident gauge. 
From the consideration of this point, it follows that no FC constraints should exist in this theory, which is consistent with the findings that all six constraints $\phi _{a}$, $\phi _{i}$, $\phi_{c}$ and $\chi$ are SC. Moreover, one can easily find that the noncommutativity of these constraints is mainly caused by the fact that the canonical momenta of $N_i$ and $\varphi$ are nonvanishing. Recalling the action \eqref{Eq: fQ ADM action2}, we find that the action contains the terms linear to the velocity of the variables $N_i$ and $\varphi$.
In the Lagrangian formulation, 
such linear terms suggest that the equation of motion only involves the first-order time derivatives of $N_i$ and $\varphi$.
Since the wave equation requires second-order time derivatives, we can interpret that $N_i$ and $\varphi$ are {\it dynamical but do not propagate}; in other words, $N^i$ and $\varphi$ carry a half d.o.f.
In Hamiltonian formulation, the six SC constraints ($\phi _{a},\phi _{i},\phi _{c},\chi$) can be regarded as determining any six phase-space variables that appear in the given constraints. 
Hence, one can choose which variables are expressed in terms of the others~\footnote{
It can be proved that with the help of Eq.~\eqref{Eq: action of CGR}, we can change the form of action~\eqref{Eq: fQ ADM action2} by applying ADM decomposition towards the action~\eqref{Eq: scalarnonmetricity} straightforwardly. In such a way the variable without time derivative in Lagrangian transform from lapse function $N$ to auxiliary field $\varphi$. However, this is only a mathematical transformation without any real physical meaning.
}. 
Finally, four of ten variables ($N, N_i, \varphi, \pi^N, \pi^i, p$) are dynamical ones, and if we choose $N_i$ and $\varphi$ as independent variables, we can again interpret that each of the variables in $N_i$ and $\varphi$ carries a half degree of freedom.

We observe that the above situation is very different from the case of GR or CGR, or even $f(R)$ and $f(T)$ gravity. 
Note that in $f(R)$ gravity, the scalar field momentum $p$ is not a constraint~\cite{Liang:2017ahj}, but rather a real dynamical variable. 
$f(R)$ gravity has $10+1=11$ apparent d.o.f.
While it preserves the diff.~symmetry, all the eight constraints (primary and secondary) are FCs, and $f(R)$ gravity has 3 physical d.o.f.~\cite{Liang:2017ahj}. 
In $f(T)$ gravity, where the building blocks are tetrad rather than metric, there are $16+1=17$ apparent d.o.f.. 
It preserves diff.~symmetry but breaks down the local Lorentz symmetry (six SC constraints), 
while they need additional 2 SC constraints to eliminate the auxiliary scalar field.
So, we have eight FC constraints and eight SC constraints, and $f(T)$ gravity has 5 physical d.o.f.~\cite{Li:2011rn}. 
We compare the three gravitational theories explicitly in Table \ref{tab:D.o.f comparison}. 

\begin{table}[htbp]\renewcommand\arraystretch{1.4}
    \caption{d.o.f comparison of geometrical trinity of gravity}
    \label{tab:D.o.f comparison}
    \centering
\begin{tabular}{|c|c|c|c|}
	\hline  & \text { Number of basic variables }  & \text { Degrees of freedom } & \text {Symmetry breaking} \\
	\hline \hline$f(R)$ & 10+1\, $(g_{\mu\nu},\phi)$ & $\left( 22-8\times 2-0 \right) /2=3$ &\makecell[c]{No symmetry is broken}\\
	\hline $f(T)$ & 16+1\, $(e^{\mu}_{\;a},\phi)$ & $\left( 34-8\times 2-8 \right) /2=5$ & \text {Local Lorentz is broken }  \\
    \hline $f(Q)$ & 10+1\, $(g_{\mu\nu},\phi)$ & $\left( 22-0-6 \right) /2=8$ & \text {Diff. is broken } \\
	\hline
	\end{tabular}
\end{table}

Although the above three theories show different features, we can also find similarities among them, giving us a better understanding of the dynamics in the $f(Q)$ gravity.
In contrast to $f(R)$ gravity, the other two theories show the SC constraints. 
As was discussed, we have eight SC constraints in $f(T)$ gravity, where six of eight reflect the breakdown of the local Lorentz symmetry, and two of eight reflect the nondynamical scalar field.
In the same way, we can interpret six SC constraints in $f(Q)$ gravity; that is, four of six reflects the breakdown of the diff.~symmetry and two of six reflect the nondynamical scalar field.
On the other hand, the symmetry is manifest, and the scalar field is dynamical in $f(R)$ gravity.
This is because the equation of motion in $f(R)$ gravity includes a fourth-order derivative, which is decomposed into two second-order derivative equations, and thus both metric and scalar fields are dynamical.

As a side remark, we would like to make speculations on what type of d.o.f. these might be. 
From viewpoint of space-time decomposition, the induced metric $h_{ij}$ contains two scalar-(S), two vector-(V), and two~tensor-(T) d.o.f. 
Due to the breakdown of diff.~symmetry, all these 6 d.o.f. are preserved. 
While the shift function $N_i$ can be decomposed into the gradient of a scalar plus a divergenceless vector, $N$ and $N_i$ contain a total of 2 scalar and 2 vector d.o.f. 
However, variables only carry a half d.o.f. as we discussed. Thus, we have in total three possible interpretations of these 8 d.o.f.; they correspond to $ (2\,\text{S}, 4\,\text{V}, 2\,\text{T})$, $(3\,\text{S}, 3\,\text{V}, 2\,\text{T})$, and $(4\,\text{S}, 2\,\text{V}, 2\,\text{T})$ respectively. 
However, determining which is the right choice is beyond the scope of our paper, and we will handle it in a future study.

\section{Conclusions and discussion} \label{sec6}
In gauge theories, redundant d.o.f. may exist, some of which are closely related to the gauge symmetry, i.e., invariance under the gauge transformation. 
In the Hamiltonian formalism, they are characterized by the presence of constraints. The symmetries inherent in a theory can be explored by performing a Hamiltonian analysis.

The three gravity theories, namely GR, TEGR, and STEGR, are well known to be equivalent to each other, despite a total derivative acting as a boundary term. Therefore, they have the same number of physical d.o.f.. 
That is why people call them the ``geometrical trinity."
However, their variants, namely $f(R)$, $f(T)$, and $f(Q)$, will not be equivalent to each other because, in a functional form, the total derivative term can no longer be a boundary term. 
While the d.o.f. of the first two theories are already known, in this paper, we developed the Hamiltonian formalism for $ f(Q) $ gravity in the framework of the coincident gauge. 

Making use of the ADM formalism and after tedious calculations, we succeeded in expressing the action in terms of ADM variables. We found that different from the widely studied $f(R)$ and $f(T)$ gravity theories, 
the ADM action of $f(Q)$ gravity contains terms linear to the velocity of its variables $N_i$ and $\varphi$, which will have significant effects on its geometric structure. These variables cannot be seen as dynamical ones because of the lack of their kinetic terms; however, their canonical momenta do not vanish as usual constraints in GR. 
Therefore, we can interpret these variables as dynamical but non-propagating d.o.f., which will also contribute to the physical degree of freedom, but lack the capability to propagate.
Then we performed the standard Dirac-Bergmann algorithm to the action and notably showed there are five primary constraints ($\phi_{a}, \phi_{i}$, $\phi_{c}$) and one secondary constraint $\chi$ in the whole system. 
After calculating the nonvanishing PBs between all constraints, we proved all of them are SC, which implies that the diff.~symmetry is broken. 
Therefore, we concluded that in four dimensions, the physical d.o.f. in general coincident relativity is 8. 
The richness of the physical d.o.f. is not only due to the broken diff.~symmetry but due to the contributions from nonpropagating variables such as $N_i$ and $\varphi$. 
We also presented a comparison of our results with $f(R)$ and $f(T)$ gravity theories. 

We further note a possible analogy to the massive gravity in the light of symmetry breaking~\cite{Hinterbichler:2011tt,Molaee:2017enn,Reyes:2022mvm}.
Massive gravity describes the self-interacting massive spin-2 field theory (so-called massive graviton), and the mass term explicitly breaks diff.~symmetry.
Consequently, we have no FC constraint, but two SC constraints show up in massive gravity.
Thus, massive gravity predicts five physical d.o.f. corresponding to the 5 helicities of the massive spin-2 field.
As seen above, we have no FC constraint but the SC constraints in the $f(Q)$ gravity with the coincident gauge. 
In the same way as the massive gravity, we can restore the diff.~symmetry by introducing the Stukelberg fields to compensate for the symmetry~\cite{Jarv:2018bgs}, which several works have argued as the covariant formulation of $f(Q)$ gravity.

The covariant formulation of $f(Q)$ gravity can complete our analysis of the mode decomposition in the present work.
Although we determined the number of physical d.o.f. and demonstrated possible interpretations for each mode,
we cannot uniquely determine the mode decomposition.
Based on the covariant formulation, we will be able to discuss the origin of each mode clearly, which provides us with the complete information of physical d.o.f.. We hope our work will reveal the dark side of exploring the basic perception of $f(Q)$ gravity and provide new insights into cosmological perturbation theory, which is also a project we plan to undertake in the future.

\begin{acknowledgments}
The authors thank M. Krššák for valuable  discussions. 
This work is supported by the National Key Research and Development Program of China under Grants No. 2021YFC2203100 and No. 2021YFA0718500, and the National Science Foundation of China under Grants No. 11875141.
\end{acknowledgments}

\bibliographystyle{apsrev4-1}
\bibliography{References.bib}

\newpage

\appendix

\section{Relation between ${}^3\mathcal{R} $ and ${}^3Q$} \label{appendix A} 
We try to determine whether there exists an analog relationship between $ {}^3\mathcal{R} $ and $ {}^3Q $ as they are in four dimensions. First, we expand the $ {}^3\mathcal{R} $ by its definition.
\begin{equation}
	\begin{aligned}
		\sqrt{-g}f'\left( \varphi \right){}^3\mathcal{R}&=\sqrt{-g}f'\left( \varphi \right)h^{ik}\mathcal{R}_{ik} \\
		&=\sqrt{-g}f'\left( \varphi \right)h^{ik}\left( \partial _l\varGamma ^l_{\ ik}-\partial _k\varGamma ^l_{\ il}+\varGamma ^l_{\ ik}\varGamma ^m_{\ lm}-\varGamma ^m_{\ il}\varGamma ^l_{\ km} \right) \\
		&=\sqrt{-g}f'\left( \varphi \right)h^{ik}\left( \partial _l\varGamma ^l_{\ ik}-\partial _k\varGamma ^l_{\ il} \right) -\sqrt{-g}f'\left( \varphi \right){}^3Q~.\label{eq. Appendix 0}
	\end{aligned}
\end{equation}
We can apply partial integral to the first two terms of right-hand side due to the integral of the action
\begin{equation}
	\begin{aligned}
		&\sqrt{-g}f'\left( \varphi \right) h^{i k} \frac{\partial \varGamma_{\ i k}^{l}}{\partial x^{l}}=\frac{\partial}{\partial x^{l}}\left(\sqrt{-g} f'\left( \varphi \right)h^{i k} \varGamma_{\ i k}^{l}\right)-\varGamma_{\ i k}^{l} \frac{\partial}{\partial x^{l}}\left(\sqrt{-g} f'\left( \varphi \right)h^{i k}\right)~, \\
		&\sqrt{-g} f'\left( \varphi \right)h^{i k} \frac{\partial \varGamma_{\ i l}^{l}}{\partial x^{k}}=\frac{\partial}{\partial x^{k}}\left(\sqrt{-g}f'\left( \varphi \right) h^{i k} \varGamma_{\ i l}^{l}\right)-\varGamma_{\ i l}^{l} \frac{\partial}{\partial x^{k}}\left(\sqrt{-g}f'\left( \varphi \right) h^{i k}\right)~.\label{eq. Appendix 1}
	\end{aligned}
\end{equation}
Making use of a useful expression
\begin{equation}
	\begin{aligned}
		& \varGamma ^m_{\ im}h^{il}-h^{ik}\varGamma ^l_{\ ik}\\
		=&\frac{1}{2}h^{mk}h^{il}\partial _ih_{mk}-h^{ik}h^{lm}( \partial _kh_{mi}-\frac{1}{2}\partial _mh_{ik} ) \\
		=&h^{mk}h^{il}( {}^3Q_{imk}-{}^3\tilde Q_{kim} )  \\
		=&{}^3{Q^{l}}-{}^3{\tilde{{Q}}^{l}}~,
	\end{aligned}
\end{equation}
we note that for the r.h.s. of the two lines in Eq.~\eqref{eq. Appendix 1}, the first terms are total derivative terms. Therefore, their difference gives rise to
\begin{equation}
	\begin{aligned}
		&  \frac{\partial}{\partial x^{l}}\left(\sqrt{-g}f'\left( \varphi \right) h^{i k} \varGamma_{\ i k}^{l}\right)-\frac{\partial}{\partial x^{k}}\left(\sqrt{-g} f'\left( \varphi \right) h^{i k} \varGamma_{\ i l}^{l}\right)\\
		=&\sqrt{h}\left[ \partial _l\left( Nf' \right) \right] \left( h^{ik}\varGamma ^l_{\ ik}-h^{il}\varGamma ^k_{\ ik} \right) +Nf'\partial _l\left[ \sqrt{h}\left( h^{ik}\varGamma ^l_{\ ik}-h^{il}\varGamma ^k_{\ ik} \right) \right] \\
		=&\sqrt{-g}f'\mathcal{D} _l[({}^3{\tilde{{Q}}^{l}}-{}^3{Q^{l}} )]+\sqrt{h}\left[\partial_k(Nf')\right]({}^3{\tilde{{Q}}^{k}}-{}^3{Q^{k}} )~,\label{eq. Appendix 2}
	\end{aligned}
\end{equation}	
while the difference of the last terms becomes
\begin{equation}
	\begin{aligned}
		& \varGamma_{\ i l}^{l} \frac{\partial}{\partial x^{k}}\left(\sqrt{-g}f'\left( \varphi \right) h^{i k}\right)-\varGamma_{\ i k}^{l} \frac{\partial}{\partial x^{l}}\left(\sqrt{-g}f'\left( \varphi \right) h^{i k}\right)\\
		=&\varGamma ^m_{\ i m}\left[ \sqrt{h}h^{ik}\partial _k(Nf')+Nf'\partial _k( \sqrt{h}h^{ik} ) \right] -\varGamma ^l_{\ i k}\left[ \partial _l\left( Nf' \right)\sqrt{h} h^{ik}+Nf'h^{ik}\partial _l\sqrt{h} +Nf'\sqrt{h}\partial_l h^{ik}\right] \\
		=&2\sqrt{-g}f'{}^3Q-\sqrt{h}({}^3{\tilde{{Q}}^{k}}-{}^3{Q^{k}})\partial_{k}(f'N)~.\label{eq. Appendix 3}
	\end{aligned}
\end{equation}	
Combining Eq.~\eqref{eq. Appendix 1}~\eqref{eq. Appendix 2}~\eqref{eq. Appendix 3}, Eq.~\eqref{eq. Appendix 0} give rise to
\begin{equation}
	\begin{aligned}
	{}^3\mathcal{R}={}^3Q-\mathcal{D} _l({}^3{Q^{l}}-{}^3{\tilde{{Q}}^{l}})~.
	\end{aligned}
\end{equation}
\section{The action \label{appendix B}}

We will compute the boundary terms of Eq.~\eqref{Eq: fQ ADM action} in detail. Integration by parts brings
\begin{equation}
  \begin{aligned}
  &\int d^{4} x \, N\sqrt{h} f'\left[\nabla_{\alpha}(Q^{\alpha}-\tilde{Q}^{\alpha})-\mathcal{D}_{l}({}^3{Q^{l}}-{}^3{\tilde{{Q}}^{l}})+2 \nabla_{\mu}\left(n^{\mu} \nabla_{\nu} n^{\nu}\right)-2 \nabla_{\nu}\left(n^{\mu} \nabla_{\mu} n^{\nu}\right)\right]\\
  =&\int d^{4} x \, f'\partial_{\alpha}\left[\sqrt{-g}(Q^{\alpha}-\tilde{Q}^{\alpha})\right]-\int d^{4} x \,  N f'\partial_{l}\left[\sqrt{h}({}^3{Q^{l}}-{}^3{\tilde{{Q}}^{l}})\right]+\int d^{4} x  \, N\sqrt{h} f'\left[2 \nabla_{\mu}\left(n^{\mu} \nabla_{\nu} n^{\nu}\right)-2 \nabla_{\nu}\left(n^{\mu} \nabla_{\mu} n^{\nu}\right)\right]\\
  =&-\int d^{4} x \, \sqrt{-g}\partial_{\alpha}f'(Q^{\alpha}-\tilde{Q}^{\alpha})+\int d^{4} x \,  \sqrt{h}\partial_{l}(N f')({}^3{Q^{l}}-{}^3{\tilde{{Q}}^{l}})+\int d^{4} x  \, N\sqrt{h} f'[2 \nabla_{\mu}\left(n^{\mu} \nabla_{\nu} n^{\nu}\right)-2 \nabla_{\nu}\left(n^{\mu} \nabla_{\mu} n^{\nu}\right)]~.\label{eq Appendix boundary}
\end{aligned}
  \end{equation}
We divide it into three parts. The second part is already completely spatial. Therefore, in the next step, we should decompose the first and third terms carefully.
  
\subsection{Decomposition of the $Q^{\alpha}-\tilde{Q}^{\alpha}$}

We apply 3+1 decomposition to the first term in~\eqref{eq Appendix boundary}
\begin{equation} - \sqrt{-g}\partial_{\alpha}f'(Q^{\alpha}-\tilde{Q}^{\alpha})= - \sqrt{-g}\partial_{0}f'(Q^{0}-\tilde{Q}^{0})- \sqrt{-g}\partial_{i}f'(Q^{i}-\tilde{Q}^{i})~, \end{equation}
which have been separated into two terms, but we have to notice that each term still contains the four-dimensional index. To clarify this, we expand them correspondingly\par

\textbf{$\bm{0-0}$ component:}
\begin{equation}
  \begin{aligned}
     Q^{0}-\tilde{Q}^{0}=&g^{0\alpha}g^{\beta \gamma}\left( \partial _{\alpha}g_{\beta \gamma}-\partial _{\beta}g_{\alpha \gamma} \right) =g^{00}g^{\beta \gamma}\left( \partial _0g_{\beta \gamma}-\partial _{\beta}g_{0\gamma} \right) +g^{0i}g^{\beta \gamma}\left( \partial _ig_{\beta \gamma}-\partial _{\beta}g_{i\gamma} \right)\\
     =&g^{00}g^{i\gamma}\left( \partial _0g_{i\gamma}-\partial _ig_{0\gamma} \right) +g^{0i}g^{0\gamma}\left( \partial _ig_{0\gamma}-\partial _0g_{i\gamma} \right) +g^{0i}g^{j\gamma}\left( \partial _ig_{j\gamma}-\partial _jg_{i\gamma} \right) \\
     =&g^{00}g^{ij}\left( \partial _0g_{ij}-\partial _ig_{0j} \right) 
     +g^{0i}g^{0j}\left( \partial _ig_{0j}-\partial _0g_{ij} \right) +g^{0i}g^{j0}\left( \partial _ig_{j0}-\partial _jg_{i0} \right) +g^{0i}g^{jk}\left( \partial _ig_{jk}-\partial _jg_{ik} \right) ~.
\end{aligned}
  \end{equation}

Inserting Eq.~\eqref{Eq: ADM metric} into above equation:
\begin{equation}
\begin{aligned} 
  Q^{0}-\tilde{Q}^{0}=&\left( -\frac{1}{N^2} \right) \left( h^{ij}-\frac{N^iN^j}{N^2} \right) \left( \dot{h}_{ij}-\partial _iN_j \right) +\frac{N^i}{N^2}\frac{N^j}{N^2}\left( \partial _iN_j-\dot{h}_{ij} \right) \\
  &+\frac{N^i}{N^2}\frac{N^j}{N^2}\left( \partial _iN_j-\partial _jN_i \right) +\frac{N^i}{N^2}\left( h^{jk}-\frac{N^jN^k}{N^2} \right) \left( \partial _ih_{jk}-\partial _jh_{ik} \right) \\
  =&-\frac{\dot{h}_{ij}h^{ij}}{N^2}+\frac{\partial ^jN_j}{N^2}+\left( \frac{N^ih^{jk}-N^jh^{ik}}{N^2} \right) \partial _ih_{jk}\\
  =&-\frac{\dot{h}_{ij}h^{ij}}{N^2}+\frac{h^{ij}}{N^2}\left( h_{kj}\partial _iN^k+N^k\partial _ih_{kj} \right) +\left( \frac{N^ih^{jk}-N^jh^{ik}}{N^2} \right) \partial _ih_{jk}\\
  =&-\frac{\dot{h}_{ij}h^{ij}}{N^2}+\frac{\partial _iN^i}{N^2}+\frac{h^{jk}N^i}{N^2}\partial _ih_{jk}~.
\end{aligned}
\end{equation}

Since the contraction of the symmetric tensor and the antisymmetric tensor is equal to zero, we eliminate some terms in the second step. Note that $h^{ij}h_{jk}=g^{i\alpha}g_{\alpha k}=\delta^{i}_{k}$  and the indices of three-dimensional tensor on hypersurface $\varSigma_t$ can be raised and lowered by the induced metric $h_{ij}$.

 \textbf{$\bm{a-a}$ component:}
\begin{equation}
  \begin{aligned} 
Q^a-\tilde{Q}^a=&g^{a\alpha}g^{\beta \gamma}\left( \partial _{\alpha}g_{\beta \gamma}-\partial _{\beta}g_{\alpha \gamma} \right) =g^{a0}g^{\beta \gamma}\left( \partial _0g_{\beta \gamma}-\partial _{\beta}g_{0\gamma} \right) +g^{ai}g^{\beta \gamma}\left( \partial _ig_{\beta \gamma}-\partial _{\beta}g_{i\gamma} \right) \\
=& g^{a0}g^{i\gamma}\left( \partial _0g_{i\gamma}-\partial _ig_{0\gamma} \right) +g^{ai}g^{0\gamma}\left( \partial _ig_{0\gamma}-\partial _0g_{i\gamma} \right) +g^{ai}g^{j\gamma}\left( \partial _ig_{j\gamma}-\partial _jg_{i\gamma} \right) \\
=&g^{a0}g^{i0}\left( \partial _0g_{i0}-\partial _ig_{00} \right) +g^{a0}g^{ij}\left( \partial _0g_{ij}-\partial _ig_{0j} \right) +g^{ai}g^{00}\left( \partial _ig_{00}-\partial _0g_{i0} \right)\\
& +g^{ai}g^{0j}\left( \partial _ig_{0j}-\partial _0g_{ij} \right) +g^{ai}g^{j0}\left( \partial _ig_{j0}-\partial _jg_{i0} \right) +g^{ai}g^{jk}\left( \partial _ig_{jk}-\partial _jg_{ik} \right)~.
\end{aligned}
\end{equation}

Inserting Eq.~\eqref{Eq: ADM metric} to above equation and after tedious calculation, it leads to
\begin{equation}
  \begin{aligned} 
    Q^a-\tilde{Q}^a
    =&\frac{N^a}{N^2}h^{ij}\left( \dot{h}_{ij}-\partial _iN_j \right)  -\frac{h^{ai}}{N^2}\partial _i\left( N_jN^j-N^2 \right) +\frac{h^{ai}}{N^2}\dot{N}_i
    +\frac{h^{ai}N^j}{N^2}\left( \partial _iN_j-\dot{h}_{ij} \right)
    +\frac{2h^{ai}N^j}{N^2} \partial _{[i}N_{j]}  \\
    &+h^{ai}h^{jk}\partial _ih_{jk}-h^{ai}h^{jk}\partial _jh_{ik}-h^{ai}\frac{N^jN^k}{N^2}\partial _ih_{jk}+h^{ai}\frac{N^jN^k}{N^2}\partial _jh_{ik}
    -\frac{2N^aN^ih^{jk}}{N^2}\partial _{[i}h_{j]k}\\
    =&\frac{N^a}{N^2}h^{ij}\dot{h}_{ij}-\frac{h^{ai}N^j}{N^2}\dot{h}_{ij}-\left(\frac{N^a}{N^2}h^{ij}\partial _iN_j+\frac{h^{ai}N^j}{N^2}\partial _jN_i\right)+2N\frac{h^{ai}}{N^2}\partial _iN+\frac{h^{ai}}{N^2}\dot{N}_i\\
    &+h^{ai}\frac{N^jN^k}{N^2}\partial _jh_{ik}-\frac{N^aN^ih^{jk}}{N^2}\partial _ih_{jk}+\frac{N^aN^ih^{jk}}{N^2}\partial _jh_{ik}+({}^3{Q^{a}}-{}^3{\tilde{{Q}}^{a}})\\
    =&\frac{N^a}{N}\left(\frac{h^{ij}\dot{h}_{ij}}{N}-\frac{N^ih^{jk}}{N}\partial _ih_{jk}- \frac{\partial _iN^i}{N}\right)-\frac{N^j}{N^2}\partial _jN^a+\frac{2\partial ^aN}{N}+\frac{\dot N^a}{N^2}+({}^3{Q^{a}}-{}^3{\tilde{{Q}}^{a}})~.
\end{aligned}
\end{equation}

On the other hand we note that
\begin{equation}
K=\nabla _{\nu}n^{\nu}=h^{ij}K_{ij}=\frac{h^{ij}}{2N}\left( \dot{h}_{ij}-\mathcal{D}_iN_j-\mathcal{D}_jN_i \right) =\frac{h^{ij}\dot{h}_{ij}}{2N}-\frac{h^{jk}N^ih_{jk,i}}{2N}-\frac{N^i_{\ ,i}}{N}~,
\end{equation}
which, upon use of the above formula, leads $0-0$ and $a-a$ components to a more condensed form
\begin{equation}
  \left\{
   \begin{aligned}
    Q^{0}-\tilde{Q^{0}} &=-\frac{2K}{N}-\frac{\partial _iN^i}{N^2}~, \\
 Q^{i}-\tilde{Q^{i}} &=\frac{N^i}{N}\left(2K+\frac{N^j{}_{,j}}{N}\right)-\frac{N^j}{N^2}\partial _jN^i+\frac{2\partial ^iN}{N}+\frac{\dot N^i}{N^2}+({}^3{Q^{i}}-{}^3{\tilde{{Q}}^{i}})~.
   \end{aligned}
   \right.
  \end{equation}

  So the 3+1 decomposition to the first part of Eq.~\eqref{eq Appendix boundary} should be
  \begin{equation} 
    \begin{aligned} 
      & \int d^{4} x \,\left[-\sqrt{-g}\partial_{\alpha}f'\left(Q^{\alpha}-\tilde{Q}^{\alpha}\right)\right]\\
    &=  \int d^{4} x \,\left[\sqrt{-g}\partial_{0}f'\left(\frac{2K}{N}+\frac{\partial _iN^i}{N^2} \right)- \sqrt{-g}\partial_{i}f'\left(\frac{2KN^i}{N}+\frac{N^i\partial_jN^j}{N^2}-\frac{N^j}{N^2}\partial _jN^i+2\partial ^i{\ln N}+\frac{\dot N^i}{N^2}+{}^3{Q^{i}}-{}^3{\tilde{{Q}}^{i}}\right)\right]~.\label{eq Appendix boundary B}
  \end{aligned}
 \end{equation}

\subsection{Decomposition of the $ \nabla_{\mu}\left(n^{\mu} \nabla_{\nu} n^{\nu}-n^{\nu} \nabla_{\nu} n^{\mu}\right)$} 

The third term of Eq.~\eqref{eq Appendix boundary} comes from 3+1 decomposition of GR, it contains two parts
\begin{equation}
  \int d^{4} x \,N\sqrt{h} f'[2 \nabla_{\mu}\left(n^{\mu} \nabla_{\nu} n^{\nu}\right)-2 \nabla_{\nu}\left(n^{\mu} \nabla_{\mu} n^{\nu}\right)]~.\label{eq: third term}
  \end{equation}

By using $ K=\nabla_{\nu}n^{\nu}$, integration by part brings the first part to

\begin{equation}
  \begin{aligned}
  \int d^{4} x \,2N\sqrt{h} f' \nabla_{\mu}\left(n^{\mu} \nabla_{\nu} n^{\nu}\right)&= \int d^{4} x \,2 f' \partial_{\mu}\left(N\sqrt{h}n^{\mu} K\right)\\
  &=-\int d^{4} x \,2 \partial_ {\mu}f'N\sqrt{h}n^{\mu} K\\
  &=-\int d^{4} x \,2 KN\sqrt{h}(\partial_ {0}f'n^{0}+\partial_ {i}f'n^{i})\\
  &=\int d^{4} x \,2 K\sqrt{h}(\partial_ {i}f'N^{i}-f^{\prime \prime} \dot{\varphi})\label{eq:term1}
\end{aligned}
\end{equation}

For the second part, $ n^{\mu} \nabla_{\mu} n_{\alpha}=a_{\alpha}$ is called “acceleration”, for an arbitrary vector $V^{\alpha}$
\begin{equation}
  \begin{aligned}
    V^{\alpha} a_{\alpha} &=V^{\alpha}n^{\mu} \nabla_{\mu} n_{\alpha}=-V^{\alpha}n^{\mu} \nabla_{\mu}\left(N \nabla_{\alpha} t\right)=-V^{\alpha}n^{\mu} \nabla_{\mu} N \nabla_{\alpha} t-V^{\alpha}N n^{\mu} \nabla_{\mu} \nabla_{\alpha} t \\
  &=\frac{V^{\alpha}}{N} n_{\alpha} n^{\mu} \nabla_{\mu} N+V^{\alpha}N n^{\mu} \nabla_{\alpha}\left(-\frac{1}{N} n_{\mu}\right)=\frac{V^{\alpha}}{N} n_{\alpha} n^{\mu} \nabla_{\mu} N+\frac{V^{\alpha}}{N} \nabla_{\alpha} N (n^{\mu} n_{\mu})-V^{\alpha}n^{\mu} \nabla_{\alpha} n_{\mu} \\
  &=\frac{V^{\alpha}}{N}\left(\nabla_{\alpha} N+n_{\alpha} n^{\mu} \nabla_{\mu} N\right)=\frac{V^{\alpha}}{N} h_{\alpha}^{\mu} \nabla_{\mu} N \\
  &=\frac{V^{\alpha}}{N} \mathcal{D}_{\alpha} N=\frac{V_{i}}{N} \mathcal{D}^{i} N=V_{i}\mathcal{D}^{i} \ln N~,
  \end{aligned}
  \end{equation}
 where we have used the torsion-free character of the connection $\nabla$ to write $\nabla_{\mu} \nabla_{\alpha} t=\nabla_{\alpha} \nabla_{\mu} t$. And the use of the normalization relation $n^{\mu} n_{\mu}=-1$ and $n^{\mu} \nabla_{\alpha} n_{\mu}=\frac{1}{2}\nabla_{\alpha}(n_{\mu}n^{\mu})=0$ in the second line. So the second part of Eq.~\eqref{eq: third term} equals

  \begin{equation}
    \begin{aligned}
    -\int d^{4} x \,2N\sqrt{h} f' \nabla_{\nu}\left(n^{\mu} \nabla_{\mu} n^{\nu}\right)&= \int d^{4} x \,2N\sqrt{h}\mathcal{D}_{j} f^{\prime} \mathcal{D}^{j} \ln N~.\label{eq:term2}
  \end{aligned}
  \end{equation}
  Combining Eqs.~\eqref{eq:term1} and~\eqref{eq:term2} we get to the decomposition of Eq.~\eqref{eq: third term}
  \begin{equation}
    \begin{aligned}
    \int d^{4} x \,N\sqrt{h} f'[2 \nabla_{\mu}\left(n^{\mu} \nabla_{\nu} n^{\nu}\right)-2 \nabla_{\nu}\left(n^{\mu} \nabla_{\mu} n^{\nu}\right)]&=\int d^{4} x \,\left[2\sqrt{h}K\left(N_{j} \mathcal{D}^{j} f^{\prime}-f^{\prime \prime} \dot{\varphi}\right)+2N\sqrt{h}\mathcal{D}_{j} f^{\prime} \mathcal{D}^{j} \ln N\right]\\
    &=\int d^{4} x \,\left[2\sqrt{h}K\left(N_{j} \partial^{j} f^{\prime}-f^{\prime \prime} \dot{\varphi}\right)+2N\sqrt{h}\partial_{i} f^{\prime} \partial^{i} \ln N\right]~.\label{eq Appendix boundary A}
  \end{aligned} \end{equation}

\subsection{The whole boundary term}

Inserting Eqs.~\eqref{eq Appendix boundary B} and Eq.~\eqref{eq Appendix boundary A} into Eq.~\eqref{eq Appendix boundary}
\begin{equation} 
  \begin{aligned} 
    &\int d^{4} x \, \left[-\sqrt{-g}\partial_{\alpha}f'\left(Q^{\alpha}-\tilde{Q}^{\alpha}\right)+ \sqrt{h}\partial_{i}(N f')({}^3{Q^{i}}-{}^3{\tilde{{Q}}^{i}})+ N\sqrt{h} f'[2 \nabla_{\mu}\left(n^{\mu} \nabla_{\nu} n^{\nu}\right)-2 \nabla_{\nu}\left(n^{\mu} \nabla_{\mu} n^{\nu}\right)]\right]\\
  &\\
  =&  \int d^{4} x \,[\sqrt{-g}\partial_{0}f'\left(\cancel{\frac{2K}{N}}+\frac{\partial _iN^i}{N^2} \right)- \sqrt{-g}\partial_{i}f'\left(\cancel{\frac{2KN^i}{N}}+\frac{N^i\partial_jN^j}{N^2}-\frac{N^j}{N^2}\partial _jN^i+\cancel{2\partial ^i{\ln N}}+\frac{\dot N^i}{N^2}\cancel{+{}^3{Q^{i}}-{}^3{\tilde{{Q}}^{i}}}\right)\\
  &+\sqrt{h}(\partial_{i}N )f'({}^3{Q^{i}}-{}^3{\tilde{{Q}}^{i}})+\cancel{\sqrt{h}N\partial_{i}( f')({}^3{Q^{i}}-{}^3{\tilde{{Q}}^{i}})}+2\sqrt{h}K\left(\cancel{N^{i} \partial_{i} f^{\prime}}-\cancel{f^{\prime \prime} \dot{\varphi}}\right)+\cancel{2N\sqrt{h}\partial_{i} f^{\prime} \partial^{i} \ln N }]\\
  &\\
  =&  \int d^{4} x \,\left[\sqrt{-g}\partial_{0}f'\frac{\partial _iN^i}{N^2} - \sqrt{-g}\partial_{i}f'\left(\frac{N^i\partial_jN^j}{N^2}-\frac{N^j}{N^2}\partial _jN^i+\frac{\dot N^i}{N^2}\right)+\sqrt{h}(\partial_{i}N )f'({}^3{Q^{i}}-{}^3{\tilde{{Q}}^{i}})\right]\\
  =&  \int d^{4} x \,\left\{\sqrt{h}\partial_{0}f'\frac{\partial _iN^i}{N} - \sqrt{h}\partial_{i}f'\left(\frac{N^i\partial_jN^j}{N}-\frac{N^j}{N}\partial _jN^i+\frac{\dot N^i}{N}\right)-N\sqrt{h}\mathcal{D}_{i}\left[f'({}^3{Q^{i}}-{}^3{\tilde{{Q}}^{i}})\right]\right\}~.
\end{aligned}
\end{equation}

The extrinsic curvature scalar $K$ in the first and third term exactly eliminate with each other, it means no time derivative of spatial metric appear in the boundary term. Finally, we reach the action~\eqref{Eq: fQ ADM action}.

\end{document}